\documentclass[aps,prx,longbibliography,twocolumn,showpacs,noeprint]{revtex4-2}

\usepackage{CJK}

\usepackage{graphicx}
\usepackage{amsmath}
\usepackage{dcolumn}
\usepackage{balance}

\usepackage{bm}
\usepackage[normalem]{ulem}
\usepackage{color}
\usepackage{hyperref}

\usepackage{epstopdf}
\newcommand{\bew}{\begin{widetext}}
\newcommand{\ew}{\end{widetext}}

\newcommand{\br}{\mathbf{r}}

\newcommand{\bff}{\mathbf{f}}

\newcommand{\tri}{\triangle}

\newcommand{\beq}{\begin{equation}}
\newcommand{\eeq}{\end{equation}}
\newcommand{\beqn}{\begin{eqnarray}}
\newcommand{\eeqn}{\end{eqnarray}}

\newcommand{\dd}{{\rm d}}

\newcommand{\la}{\langle}

\newcommand{\ra}{\rangle}

\newcommand{\vnab}{{\bf \nabla}}

\allowdisplaybreaks[1]
\usepackage{amsfonts}
\usepackage{placeins}
\usepackage{float}
\usepackage[percent]{overpic}

\begin{document}
\title{Vertex Model Mechanics Explain the Emergence of Centroidal Voronoi Tiling in Epithelia}
	\author{Sulaimaan Lim}
	\affiliation{Department of Bioengineering, Imperial College London, South Kensington Campus, London SW7 2AZ, U.K.}
	\author{Julien Vermot}	
	\email{j.vermot@imperial.ac.uk}
	\affiliation{Department of Bioengineering, Imperial College London, South Kensington Campus, London SW7 2AZ, U.K.}
	\author{Chiu Fan Lee}
	\email{c.lee@imperial.ac.uk}
	\affiliation{Department of Bioengineering, Imperial College London, South Kensington Campus, London SW7 2AZ, U.K.}
	\begin{abstract}
  Epithelia are confluent cell layers that self‑organize into polygonal networks whose geometry encodes their mechanical state. A principal driver is the tunable contractility of the actomyosin cortex, which links cell‑junction tension to tissue architecture. Notably, epithelial tilings frequently resemble centroidal Voronoi tessellations (CVTs), yet the physical origin of this resemblance has remained unclear. Here, using a minimal vertex model that relates cell shape to a mechanical energy, we show that CVT‑like patterns arise naturally in the solid (rigid) regime of tissues. Analytical theory reveals that isotropic strain minimization drives cell centroids toward Voronoi configurations, a result we corroborate with a analytical mean‑field formulation of the vertex model. We further demonstrate that physiologically relevant perturbations—such as cyclic stretch—shift tissues into distinct, geometrically disordered CVT states, and that these shifts provide quantitative, image‑based readouts of mechanical state. Together, our results identify a mechanical origin for CVT‑like organization in epithelia and establish a geometric framework that infers tissue stresses directly from morphology, offering broadly applicable metrics for assessing rigidity and remodeling in living tissues.

	\end{abstract}
		
\maketitle


\begin{figure}[t]
    \centering
    \includegraphics[width=\linewidth]{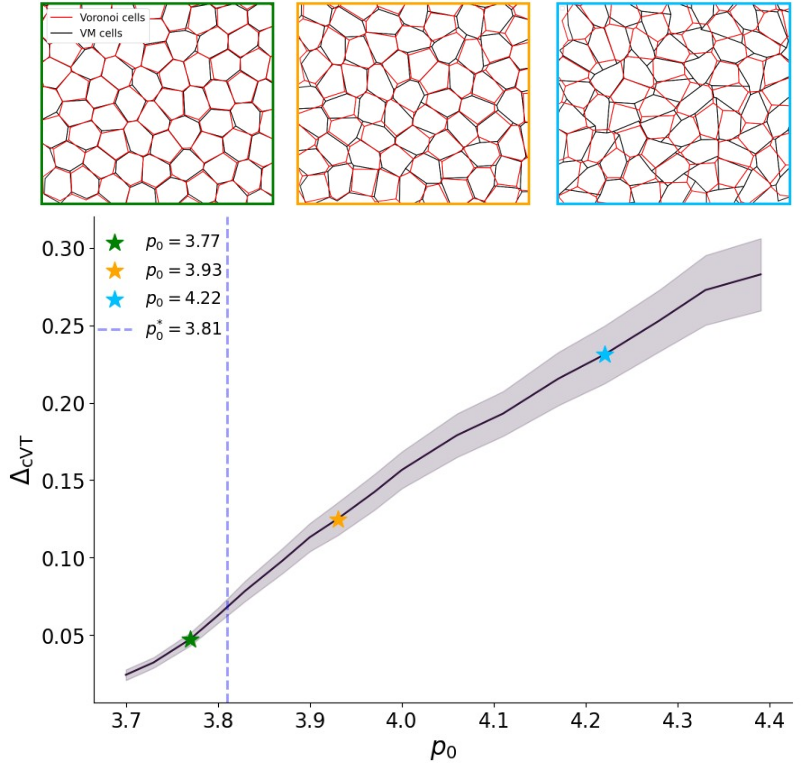}
    \caption{\textit{Vertex models (VM) approach CVTs as the shape index $p_0$ decreases.} 
The VM--CVT deviation metric $\triangle_{\mathrm{CVT}}$, which measures the average 
dimensionless difference between a VM tessellation and its corresponding 
centroid-seeded Voronoi diagram, decreases monotonically with the shape index $p_0$. 
Because lowering $p_0$ generally increases the mechanical stiffness of VM configurations, 
stiffer states exhibit patterns increasingly similar to centroidal Voronoi tessellations (CVTs). 
Representative VM networks (black) and their closest CVTs (red) at selected values of $p_0$ are 
shown in the top-row insets.
Cells in the solid phase---i.e., for $p_0$ below the rigidity transition at 
$p_0 \simeq 3.81$ (blue dashed line)---produce near-perfect CVT-like patterns. 
Further decreasing $p_0$ deepens the solid regime and yields increasingly isotropic 
arrangements, with $p_0 = 3.72$ producing a nearly ideal honeycomb lattice. 
The typical rigidity transition at $p_0 = 3.81$ is indicated by the blue dotted line; 
notably, $\triangle_{\mathrm{CVT}}$ varies smoothly across this transition.
}
    \label{fig1}
\end{figure}

\section{Introduction}
Planar confluent tissues are assemblies of cells packed tightly together without gaps, forming a continuous two-dimensional sheet. Such tissues arise both in vivo and in vitro, spanning diverse biological systems---from early embryonic layers to the endothelial linings of blood vessels \cite{halpernRoleIntestinalEpithelial2015,wallezEndothelialAdherensTight2008}. Epithelia, the most common class of planar confluent tissues, establish cohesive barriers that maintain homeostasis and undergo extensive remodelling during development and wound healing \cite{bruguesForcesDrivingEpithelial2014,julicherEmergenceTissueShape2017}---processes that critically depend on the generation and regulation of mechanical forces. The principal architectures through which cells generate these forces are closely tied to cell shape, relying on structures such as the actomyosin cortex and other cytoskeletal filament networks \cite{salbreuxActinCortexMechanics2012,khalilgharibiStressRelaxationEpithelial2019a,duqueRuptureStrengthLiving2024}. Consequently, the geometric organization, or tiling, of epithelial cells is directly coupled to the underlying biomechanical stress patterns within the tissue \cite{kongExperimentalValidationForce2019a,braunsGeometricBasisEpithelial2024,kimCellShapesPatterns2015a}.

Recent advances in high-resolution imaging have made it possible to visualize individual cell shapes within confluent tissues. Combined with the development of tools that allow localized perturbations of force generation \cite{guckOpticalStretcherNovel2001,serwaneVivoQuantificationSpatially2017}, these techniques have spurred extensive research on cell configurations in confluent tissues and the mechanical models that describe them. Particular attention has been given to the distribution of cellular aspect ratios or shape indices, which has been shown to follow a universal form across diverse types of confluent tissues \cite{atiaGeometricConstraintsEpithelial2018,sanchez-gutierrezFundamentalPhysicalCellular2016,sadhukhanOriginUniversalCell2022a,li2GammaDistributions2021}. In the mechanical vertex model \cite{nagaiDynamicCellModel2001,farhadifarInfluenceCellMechanics2007}, key parameters of this universal distribution have been linked to the rigidity of the cell layer \cite{biDensityindependentRigidityTransition2015}, with experimental studies demonstrating that a decrease in rigidity is accompanied by an increase in the shape index \cite{parkUnjammingCellShape2015}.

Beyond cellular aspect ratios, the organization of the epithelial junctional network has also received considerable attention. Although this network varies between individual samples, conserved geometrical properties have been identified that relate epithelial tilings to Voronoi tessellations, particularly to specialized forms known as centroidal Voronoi tessellations (CVTs) \cite{Voronoi1908}. In a pioneering 1978 study \cite{hondaDescriptionCellularPatterns1978}, Honda first observed that many cultured epithelial tissues exhibit distinctive CVT-like patterns. A Voronoi tessellation partitions space into regions according to proximity to a set of seed points, such that each region contains all points closer to its seed than to any other. These regions are necessarily convex polygons. A tessellation becomes a CVT when each seed point coincides with the centroid (center of mass) of its corresponding region.

Despite these observations, the mechanisms underlying the frequent resemblance between epithelial tilings and centroidal Voronoi tessellations (CVTs) remain poorly understood. In this work, we address this question using the vertex model (VM)—a standard computational framework for studying planar confluent tissues. Combining simulation and analytical approaches, we establish direct links between vertex model states, cell shape indices, and CVTs, revealing a mechanical basis for CVT-like organization in epithelial layers. We further demonstrate how this relationship is modulated under physiologically relevant perturbations, such as tissue stretching or pulsatile deformation \cite{gudipatyMechanicalStretchTriggers2017,rappelSpatiotemporalControlWave1999}. Finally, we leverage these insights to develop a method for inferring mechanical stretch in confluent tissues, as represented in the VM, using junctional network information alone.

\section{Vertex models \& centroidal Voronoi tessellations.}
The Vertex model is a cell-level computational model of confluent tissues that has become a standard tool in modelling cell dynamics and configuration in a confluent tissue. By construction, all cells are polygonal in shape and are thus determined by their respective vertices. The model energy of an $N$-cell system is given by
\beq
E_{\rm VM} = \sum_{i=1}^N \left[K_A (A_i-A_0)^2 + K_P(P_i-P_0)^2\right]\ ,
\eeq
where $A_i$ denotes the area of the $i$-th cell, $A_0$ the preferred cell area, $P_i$ the perimeter of $i$-th cell, $P_0$ the preferred perimeter, and $K_A$ and $K_P$ are the ``stiffness coefficients" for the area and perimeter terms, respectively. The dynamics of the system is then typically modelled by temporally evolving the position of the vertices based on the derivatives of the tissue energy as follows:
\beq
\frac{\dd \br_{ij}}{\dd t}= \eta \vnab_{\br_{ij}} E_{\rm VM} + \bff
 \ ,
\eeq
where $\br_{ij}$ denotes the position of the $j$-th vertex of the $i$-th cell,  $\eta$ is the damping coefficient and $\bff$ is a Gaussian noise term with zero means and correlation as follows:
\beq
\la \bff (\br, t) \bff(\br',t')\ra= 2D \delta^2(\br-\br') \delta(t-t') 
 \ .
\eeq
A key emergent property of the VM is that a ``solid-to-fluid" transition occurs at the particular value of the cell shape index, $p$, defined as the ratio between the cell perimeter and the square root of the cell area, i.e., $p_i = P_i/\sqrt{A_i}$. Specifically, when $p_0 \equiv P_0/\sqrt{A_0}$ is below the critical threshold $p_c \simeq 3.81$, the tissue behaves like a solid (e.g., shear stress sustaining), and when $p_0 >p_c$, the tissue behaves like a fluid (e.g., flowing under shear). Furthermore, for values of $p_0 \leq 3.72$, the ground state of the vertex model is well defined as a hexagonal (honeycomb) lattice, which is by definition a CVT.

We now establish a connection between the VM cellular geometry and CVTs. To this end, we introduce a simple metric, denoted by $\tri_{\rm CVT}$, which quantifies the deviation of a VM cellular network at varying values of $p_0$ from a perfect CVT pattern, based on the displacements of the corresponding vertices in the two networks [see supplemental material ({\bf SM})]. For $N=256$ cells with 2D periodic boundary conditions we generate initial states of varying of $3.7\leq p_0 \leq 4.5$ following the simulation protocol in Killeen {\it{et al.}} \cite{killeenPolarFluctuationsLead2021}.

Strikingly, Fig.~1 reveals a clear correspondence between CVT structure and VM states: as $p_0$ decreases, VM configurations vary smoothly and approach near-perfect CVT organization towards $p_0 \approx 3.72$, at which point the vertex model can satisfy a honeycomb lattice. Decreasing $p_0$ relates to an increase in structural homogeneity (Fig S1) \cite{biDensityindependentRigidityTransition2015}, and whilst there is no explicit link between isotropy and CVTs, most CVT seeking algorithms approach CVTs with associated increases in structural homogeneity \cite{sanchez-gutierrezFundamentalPhysicalCellular2016,srinivasanStochasticVoronoiTessellations2024}. Although cells are most CVT-like in the solid phase, no sharp transition is observed at the solid–fluid boundary ($p_0 \approx 3.81$) with respect to $\tri_{\rm CVT}$.

\begin{figure*}[t]
    \centering
    \includegraphics[width=\textwidth]{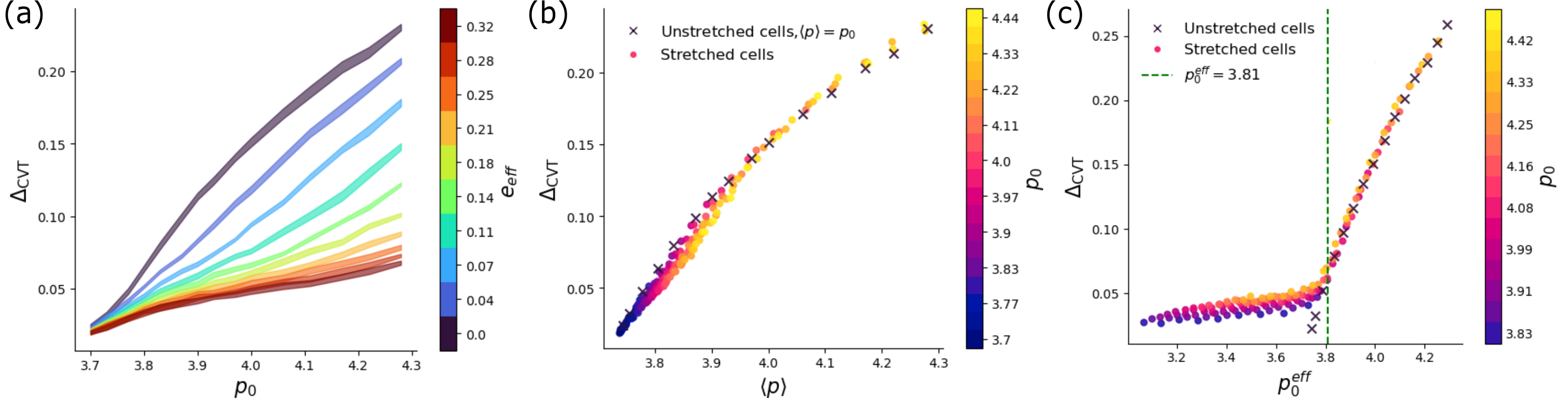}

    \caption{\textit{Oscillatory stretching drives fluid cells toward CVT patterns by lowering the average shape index $\la p\ra$.}(a) The VM--CVT deviation $\triangle_{\mathrm{CVT}}$ as a function of $p_0$ under varying fractional stretch $e_{\mathrm{eff}}$. Increasing stretch systematically reduces $\triangle_{\mathrm{CVT}}$ at fixed $p_0$. For oscillatory stretching with maximum amplitude $e$, the effective stretch is taken as the time-averaged value $e_{\mathrm{eff}} = e / 2$. (b) Replotting all simulation data for varying $p_0$ and $e$ (Fig.\ 2a) in terms of the mean cell shape $\langle p \rangle$, with colors indicating the corresponding $p_0$. The collapse of these curves onto the unstretched reference curve (black) indicates that stretch drives cells toward CVT-like configurations by effectively reducing the target shape index $p_{0}^{\mathrm{eff}}$. (c) Replotting all simulation data directly in terms of $p_0^{eff}=\frac{p_0}{1+e}$ we show a good curve collapse in the fluid phase, showing that the reduction of $\tri_{\rm CVT}$ is directly due to the stretch-induced rescaling of $p_0^{eff}$. 
    }
    \label{fig2}
\end{figure*}

\section{ Analytical connection between the VM and CVTs}
To establish an analytical connection between vertex model (VM) configurations and centroidal Voronoi tessellations (CVTs) in the low-$p_0$ limit, we adopt a single cell mean-field approximation of the vertex model energy as described by Huang {\it{et al.}} \cite{huangShearDrivenSolidificationNonlinear2022}. The key idea is to relate the VM cell energy to the \textit{quantizer energy} $E_q$, which quantifies the spatial distribution of cell centroids within a confluent tessellation \cite{CentroidalVoronoiTessellations}.  

The quantizer energy is defined as
\begin{equation}
E_q = \sum_{i=1}^{n} \int_{R_i} \| \mathbf{x} - \mathbf{x}_i \|^2 \, d\mathbf{x},
\end{equation}
where each $R_i$ denotes the region associated with a cell centroid $\mathbf{x}_i$. This energy is minimized when the regions $\{R_i\}$ form a Voronoi tessellation of the points $\{\mathbf{x}_i\}$, and each $\mathbf{x}_i$ coincides with the centroid of its region. Hence, centroidal Voronoi tessellations locally minimise $E_q$.

To connect this construction to the vertex model, we approximate each cell as a small, area-preserving affine deformation of a regular polygon. 

The deformation tensor
\begin{equation}
D =
\begin{pmatrix}
d_{xx} & 0 \\
d_{xy} & 1/d_{xx}
\end{pmatrix}
\end{equation}
maps the undeformed vertex positions to their new locations. Under this deformation, the mean-field VM energy of an $n$-sided cell is
\begin{equation}
E_{\text{cell}} = \frac{1}{2}\alpha t\, m(D)^2 + \frac{1}{4}\beta m(D)^4,
\end{equation}
where \( m^2(D) = \tfrac{2}{5}(\|D\|^2 - 2) \) to quadratic order (see SM). Here, $\alpha$ and $\beta$ are positive constants proportional to $p_0^2$, and $t \propto (p_c - p_0)$ determines the phase: $t > 0$ in the solid regime and $t < 0$ in the fluid regime.  

For $t > 0$, the energy exhibits a single gapped minimum at $m = 0$, corresponding to a regular $n$-gon. For $t < 0$, a finite-strain minimum appears, reflecting fluid-like behaviour and the loss of rigidity. In the solid phase ($t > 0$), the leading-order expansion
\begin{equation}
E_{\text{cell}} \propto p_0^{2}(\|D\|^2 - 2)
\end{equation}
describes elastic deformations of nearly regular polygons, consistent with the known elastic behaviour of the VM.

\begin{figure*}[t]
    \centering
    \includegraphics[width=\textwidth]{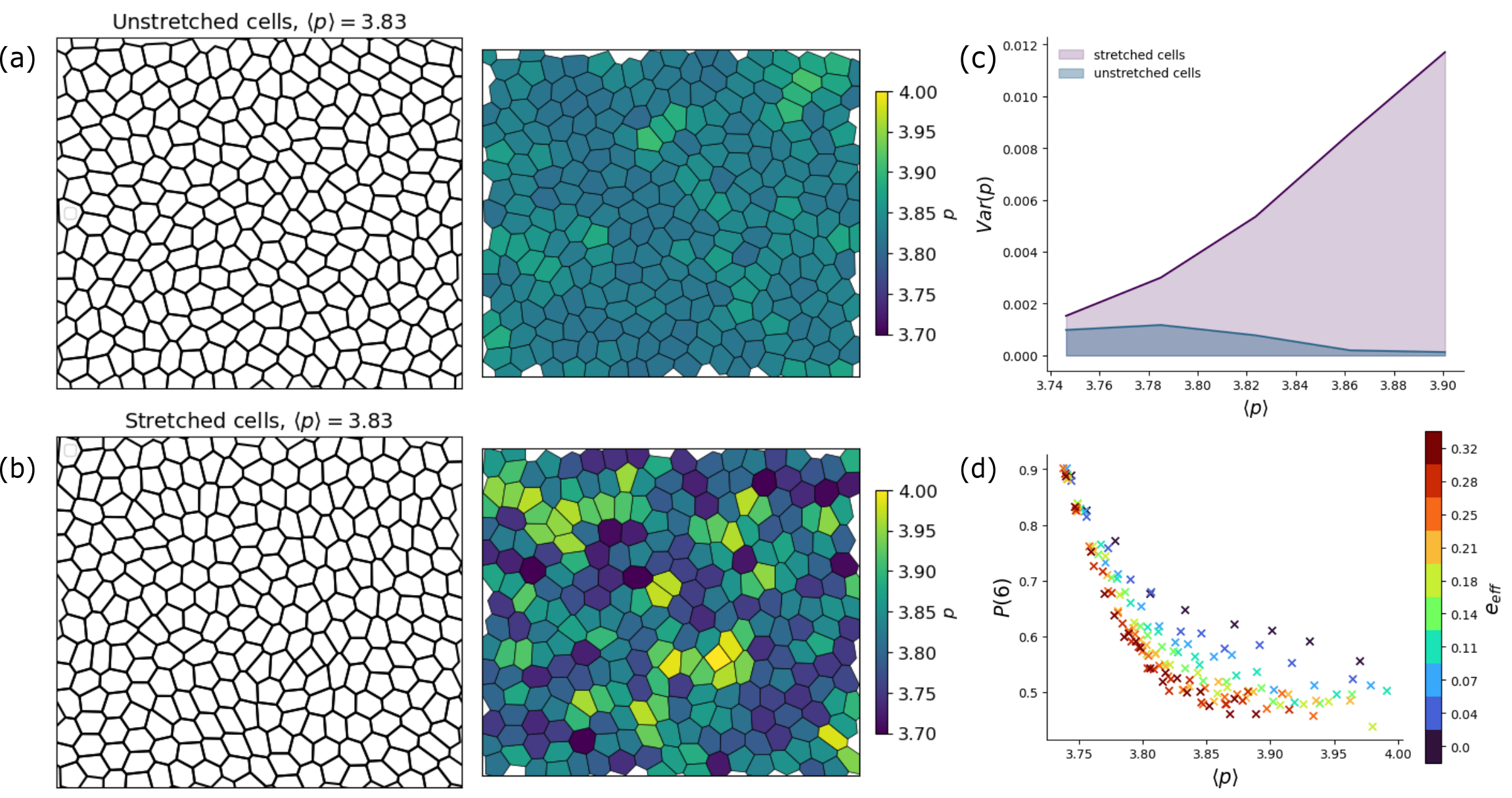}

    \caption{\textit{Distinguishing unstretched and stretched CVT-like states at constant $\langle p \rangle$ via persistent structural disorder.}
    Representative unstretched (a) and stretched (b) vertex-model tissues at fixed $\langle p \rangle = 3.83$ 
    are shown, with the VM tessellation (left) in black and the corresponding cell shape indices color-coded (right).
    (c) Stretched tissues exhibit a systematically larger variance in $\langle p \rangle$ than unstretched tissues, 
    across all simulations with $3.7 < p_0 < 4.5$ and $0 < e < 0.5$.(d) At fixed $\langle p \rangle$, the proportion of hexagons $P(6)$ decreases with increasing stretch, 
    reflecting persistent structural disorder inherited from the fluid-like initial state.Although unstretched and stretched tissues appear similar when compared using average observables (Fig.~\ref{fig2}b), they are clearly distinguishable through their cell-to-cell variability (c) 
    and monolayer polygonal structure (d). Because both $P(6)$ and $\langle p \rangle$ are purely observable quantities, they provide an experimentally 
    accessible readout of the underlying applied stretch.}
    \label{fig3}
\end{figure*}

Following work by Li et al., \cite{li2GammaDistributions2021}, the moment of inertia $I_i$ of each polygonal region under a small, area-preserving affine deformation satisfies
\begin{equation}
I_i \approx m_0 \|D\|^2,
\end{equation}
where $m_0$ is the moment of inertia of the undeformed regular $n$-gon. Substituting this into the definition of $E_q$ yields
\begin{equation}
E_q = \sum_{i=1}^{n} m_0 \|D\|^2
    = \sum_{i=1}^{n} \frac{m_0}{p_0^2} E_{\text{cell}} + \text{const.}
\end{equation}
Thus, up to a constant factor, the quantizer energy is proportional to the total vertex model energy in the solid phase:
\begin{equation}
E_q \propto E_{\mathrm{VM}}.
\end{equation}
This equivalence implies that CVT configurations locally minimize the VM energy.

The above mean-field analysis applies most directly to regular planar lattices, such as the hexagonal lattice. However, the ratio $\tfrac{m_0}{p_0^2}$ varies only weakly, within 3 percent, for realistic polygons with $5 \leq n \leq 7$ (SM), indicating that the equivalence holds approximately for mixed-polygon tilings as well.

In summary, within the solid phase, vertex-model cells behave as elastically deformed regular polygons, and CVT configurations correspond to states that locally minimize the collective mechanical energy of the tissue. This provides an analytical explanation for the observed correspondence between VM equilibria and CVT geometry.

\section{Effect of stretching on the VM--CVT connection}
Having established the equivalence between VM configurations and CVTs at low $p_0$, we next examine how external stretching modifies this correspondence. Stretch-induced remodelling plays a central role in epithelial mechanics, and we therefore ask whether physiologically relevant perturbations, such as isotropic or oscillatory stretch, can drive initially non-CVT configurations toward CVT-like organization.  

Isotropic stretching of relaxed VM cells increases their areas beyond $A_0$, effectively reducing the target shape index to
\[
p^{\mathrm{eff}}_0 = \frac{p_0}{1+e},
\]
for a fractional isotropic strain $e$ (i.e., scaling all coordinates by $1+e$) \cite{hernandezFiniteElasticityVertex2023,nestor-bergmannRelatingCellShape2018}. Because the rigidity transition in the VM occurs at $\langle p \rangle \approx 3.81$, stretching shifts the system toward lower $p^{\mathrm{eff}}_0$, thus stiffening the tissue and potentially inducing a fluid-to-solid transition. Therefore, stretch can act as a mechanical control parameter that restores order and promotes CVT-like geometry.

In the mean-field picture, fluid-phase cells with high $p_0$ minimize their energy at larger values of $\|D\|$, within a degenerate, low-rigidity regime \cite{huangShearDrivenSolidificationNonlinear2022}. Reducing $p^{\mathrm{eff}}_0$ through stretch can drive the system into the solid phase, where the cell's energy is proportional to $\|D\|^2$and the $E_{VM} \propto E_q$ equivalence applies. Consequently the tissue moves from a high quantizer-energy ($E_q$) configuration to a local $E_q$ minimum, corresponding to a CVT state.  

We tested this prediction in simulations of homogeneous VM tissues ($3.7 \le p_0 \le 4.5$) subjected to isotropic oscillatory stretch by a factor $(1 + e)$, with $0 \le e \le 0.5$ (exact simulation details in SM). We choose an oscillatory frequency of $\sim 2$ Hz, inspired by the developmental zebrafish heartbeat \cite{delucaZebraBeatFlexiblePlatform2014}, and a relaxation timescale of $\sim$ 1 minute previously estimated for developmental zebrafish KV and tailbud cells \cite{erdemci-tandoganTissueFlowInduces2018b,serwaneVivoQuantificationSpatially2017}.
As predicted, increasing $e$ monotonically decreased the deviation measure $\tri_{\mathrm{CVT}}$, indicating that stretch promotes CVT-like order even in initially fluid configurations (Fig.~\ref{fig2}a). 
To compare stretched and unstretched conditions, we replotted $\tri_{\mathrm{CVT}}$ against the observed shape index $\langle p \rangle$, which for equilibrated systems should resemble $p^{\mathrm{eff}}_0$ for values above $3.72$---a honeycomb lattice (Fig.~\ref{fig2}b). The results (Fig.~\ref{fig2}a) collapse onto the same curve as unstretched homogeneous tissues Fig.~\ref{fig1}, suggesting that $\tri_{\mathrm{CVT}}$ depends primarily on $p^{\mathrm{eff}}_0$, independent of the specific values of $p_0$ or $e$. We then demonstrate the functional dependence of $\tri_{\mathrm{CVT}}$ on $p^{\mathrm{eff}}_0$ explicitly (Fig.~\ref{fig2}c), assuming for oscillatory stretching a time averaged effective shape index,
 $p^{\mathrm{eff}}_0=\langle p^{\mathrm{ eff}}_0(t) \rangle=\frac{1}{1+\langle e(t) \rangle }$. In the effective fluid phase, $p^{\mathrm{eff}}_0> 3.81$ we see a clear collapse of all $\tri_{\mathrm{CVT}}$ measures to a singular $p^{\mathrm{eff}}_0$ curve. Below $ p^{\mathrm{eff}}_0 \approx 3.81$, where the equivalence $E_{\mathrm{VM}} \propto E_q$ holds and cells are driven to local CVT states, there is
  deviation of the stretched and unstretched curves dependent on the initial conditions. Below $ p^{\mathrm{eff}}_0 \approx 3.72$, the honeycomb limit at which cells can geometrically satisfy $p=p_0$, there is a slow decrease in $\tri_{\mathrm{CVT}}$ with decreasing $p^{\mathrm{eff}}_0$, due an increase in energy of solid phase states ($t \propto (p_c - p_0)$), which increases the depth of local minima in the quantiser energy landscape. The collapse of data with the time-averaged effective shape-index suggests cyclic-stretching is felt as a time-averaged static stretch, due to the separation of timescales between stretching frequency and the vertex-model relaxation timescale, and we by directly simulating static stretch (Fig.~S3 in SM).

Thus, stretching drives cells toward CVT configurations by effectively lowering $p^{\mathrm{eff}}_0$, rather than altering other model parameters. Notably, tissues near or within the solid phase ($\langle p \rangle \lesssim 3.9$) show slightly reduced $\tri_{\mathrm{CVT}}$, indicating enhanced geometric regularity.

\section{Structural consequences of stretching}
Having shown that stretching drives cells toward CVT configurations as effectively, or even more so, than homogeneous simulations at equivalent mean shape indices $\langle p \rangle$, and having established an analytical basis for this behaviour, we now examine how stretching modifies the detailed structural organization of these CVT-like states.  

Fig.~\ref{fig3}a-c shows that for states with $\langle p \rangle \leq 3.9$, stretched cells exhibit higher variance in shape indices compared to unstretched counterparts. To further quantify these structural differences, we analyzed the hexagon fraction $P(6)$---the proportion of six-sided cells, a standard descriptor of epithelial organization known to vary with mechanical conditions \cite{farhadifarInfluenceCellMechanics2007,sanchez-gutierrezFundamentalPhysicalCellular2016}. Organizing our simulations by the observable pair $(\langle p \rangle, P(6))$ (Fig.~\ref{fig3}d), we find that this pair is uniquely determined by $(p_0, e)$. For a fixed $\langle p \rangle$, the hexagonal fraction $P(6)$ decreases with increasing stretch $e$ up to a threshold level, indicating that stretching promotes CVT-like geometry whilst maintaining structural disorder.
  
These differences arise naturally from the energy equivalence $E_{\mathrm{VM}} \propto E_q$: strong stretch ($p_0^{\mathrm{eff}} < 3.81$) rapidly quenches the system into local minima of $E_q$, yielding low $\tri_{\mathrm{CVT}}$ but higher geometric variance. In contrast, unstretched cells, which relax more gradually, reach shallower minima (Fig.~S2 in SM) and more homogeneous configurations. Importantly, these distinctions are detectable directly from observable quantities in VM simulations, suggesting that similar structural signatures could, in principle, be used to infer tissue stretch from imaging data alone.

\section{Discussion \& Summary}
We have shown that a simple, homogeneous vertex model (VM) naturally generates centroidal Voronoi tessellation (CVT)–like patterns near the solid phase, 
and that proximity to CVTs is directly controlled by the key vertex-model parameter $p_0$. 
This behaviour follows from an analytical equivalence between the VM energy and the quantizer energy, extending previous theoretical work 
\cite{huangShearDrivenSolidificationNonlinear2022,li2GammaDistributions2021}. 
We further demonstrated that external stretch can drive fluid-like tissues toward CVT organization by effectively reducing the target shape index $p_0$.

Despite having identical mean shape indices $\langle p \rangle$, stretched and unstretched CVT-like states remain structurally distinct: 
stretch acts as a rapid quench in the quantizer energy $E_q$, trapping tissues in local minima characterized by increased variance in cell shape indices 
and a reduced fraction of hexagons. These quantities therefore serve as experimentally accessible signatures of underlying mechanical stretch, 
enabling the influence of external forces on epithelia to be inferred directly from imaging data. 
In addition to static deformation, fast oscillatory stretch also drives tissues toward disordered CVT-like states due to a separation of timescales 
between junctional relaxation and the imposed stretch frequency, a regime that encompasses physiologically relevant conditions in the embryonic zebrafish.

The discriminating power of structural disorder in stretched epithelia parallels findings by Wang \emph{et al.} \cite{wangAnisotropyLinksCell2020}, 
who showed that nematic alignment and shape index jointly determine whether sheared tissues undergo T1 transitions and thus behave as solids or fluids. 
Similarly, isotropic stretch effectively quenches the target shape index $p_{0}^{\mathrm{eff}}$, 
and direct reductions of $p_0$ in homogeneous simulations recapitulate the behaviour observed under stretch (Fig.~{S3} in SM). 
Recent studies have emphasized the importance of rigidity transitions during development 
\cite{petridouRigidityPercolationUncovers2021,mongeraFluidtosolidJammingTransition2018,shellardSculptingStiffnessRigidity2023}; 
our results suggest that structural disorder in CVT-like epithelia—linked to the degree of quenching in $p_0$—may provide a morphological readout 
of past rigidification events in vivo.

In conclusion, our findings establish a direct mechanical correspondence between the VM and CVT frameworks. 
Stretch acts as a tunable control parameter that quenches tissues into CVT-like states with distinct geometric signatures, 
offering a potential diagnostic for identifying mechanically constrained or rigidified epithelia. 
Moreover, the emergence of CVT-like organization itself may help assess the applicability of vertex-model mechanics, 
much as the satisfaction of Delaunay triangulation constraints supports the use of force-balance models \cite{braunsGeometricBasisEpithelial2024}. 
Together, these results highlight the utility of geometric measures—particularly CVT similarity—in linking cellular organization, 
mechanical state, and tissue-scale morphogenesis.


\begin{thebibliography}{36}%
\makeatletter
\providecommand \@ifxundefined [1]{%
 \@ifx{#1\undefined}
}%
\providecommand \@ifnum [1]{%
 \ifnum #1\expandafter \@firstoftwo
 \else \expandafter \@secondoftwo
 \fi
}%
\providecommand \@ifx [1]{%
 \ifx #1\expandafter \@firstoftwo
 \else \expandafter \@secondoftwo
 \fi
}%
\providecommand \natexlab [1]{#1}%
\providecommand \enquote  [1]{``#1''}%
\providecommand \bibnamefont  [1]{#1}%
\providecommand \bibfnamefont [1]{#1}%
\providecommand \citenamefont [1]{#1}%
\providecommand \href@noop [0]{\@secondoftwo}%
\providecommand \href [0]{\begingroup \@sanitize@url \@href}%
\providecommand \@href[1]{\@@startlink{#1}\@@href}%
\providecommand \@@href[1]{\endgroup#1\@@endlink}%
\providecommand \@sanitize@url [0]{\catcode `\\12\catcode `\$12\catcode
  `\&12\catcode `\#12\catcode `\^12\catcode `\_12\catcode `\%12\relax}%
\providecommand \@@startlink[1]{}%
\providecommand \@@endlink[0]{}%
\providecommand \url  [0]{\begingroup\@sanitize@url \@url }%
\providecommand \@url [1]{\endgroup\@href {#1}{\urlprefix }}%
\providecommand \urlprefix  [0]{URL }%
\providecommand \Eprint [0]{\href }%
\providecommand \doibase [0]{https://doi.org/}%
\providecommand \selectlanguage [0]{\@gobble}%
\providecommand \bibinfo  [0]{\@secondoftwo}%
\providecommand \bibfield  [0]{\@secondoftwo}%
\providecommand \translation [1]{[#1]}%
\providecommand \BibitemOpen [0]{}%
\providecommand \bibitemStop [0]{}%
\providecommand \bibitemNoStop [0]{.\EOS\space}%
\providecommand \EOS [0]{\spacefactor3000\relax}%
\providecommand \BibitemShut  [1]{\csname bibitem#1\endcsname}%
\let\auto@bib@innerbib\@empty
\bibitem [{\citenamefont {Halpern}\ and\ \citenamefont
  {Denning}(2015)}]{halpernRoleIntestinalEpithelial2015}%
  \BibitemOpen
  \bibfield  {author} {\bibinfo {author} {\bibfnamefont {M.~D.}\ \bibnamefont
  {Halpern}}\ and\ \bibinfo {author} {\bibfnamefont {P.~W.}\ \bibnamefont
  {Denning}},\ }\bibfield  {title} {\bibinfo {title} {The role of intestinal
  epithelial barrier function in the development of {{NEC}}},\ }\href
  {https://doi.org/10.1080/21688370.2014.1000707} {\bibfield  {journal}
  {\bibinfo  {journal} {Tissue Barriers}\ }\textbf {\bibinfo {volume} {3}},\
  \bibinfo {pages} {e1000707} (\bibinfo {year} {2015})}\BibitemShut {NoStop}%
\bibitem [{\citenamefont {Wallez}\ and\ \citenamefont
  {Huber}(2008)}]{wallezEndothelialAdherensTight2008}%
  \BibitemOpen
  \bibfield  {author} {\bibinfo {author} {\bibfnamefont {Y.}~\bibnamefont
  {Wallez}}\ and\ \bibinfo {author} {\bibfnamefont {P.}~\bibnamefont {Huber}},\
  }\bibfield  {title} {\bibinfo {title} {Endothelial adherens and tight
  junctions in vascular homeostasis, inflammation and angiogenesis},\ }\href
  {https://doi.org/10.1016/j.bbamem.2007.09.003} {\bibfield  {journal}
  {\bibinfo  {journal} {Biochimica et Biophysica Acta (BBA) - Biomembranes}\
  }\textbf {\bibinfo {volume} {1778}},\ \bibinfo {pages} {794} (\bibinfo {year}
  {2008})}\BibitemShut {NoStop}%
\bibitem [{\citenamefont {Brugués}\ \emph {et~al.}(2014)\citenamefont
  {Brugués}, \citenamefont {Anon}, \citenamefont {Conte}, \citenamefont
  {Veldhuis}, \citenamefont {Gupta}, \citenamefont {Colombelli}, \citenamefont
  {Muñoz}, \citenamefont {Brodland}, \citenamefont {Ladoux},\ and\
  \citenamefont {Trepat}}]{bruguesForcesDrivingEpithelial2014}%
  \BibitemOpen
  \bibfield  {author} {\bibinfo {author} {\bibfnamefont {A.}~\bibnamefont
  {Brugués}}, \bibinfo {author} {\bibfnamefont {E.}~\bibnamefont {Anon}},
  \bibinfo {author} {\bibfnamefont {V.}~\bibnamefont {Conte}}, \bibinfo
  {author} {\bibfnamefont {J.~H.}\ \bibnamefont {Veldhuis}}, \bibinfo {author}
  {\bibfnamefont {M.}~\bibnamefont {Gupta}}, \bibinfo {author} {\bibfnamefont
  {J.}~\bibnamefont {Colombelli}}, \bibinfo {author} {\bibfnamefont {J.~J.}\
  \bibnamefont {Muñoz}}, \bibinfo {author} {\bibfnamefont {G.~W.}\
  \bibnamefont {Brodland}}, \bibinfo {author} {\bibfnamefont {B.}~\bibnamefont
  {Ladoux}},\ and\ \bibinfo {author} {\bibfnamefont {X.}~\bibnamefont
  {Trepat}},\ }\bibfield  {title} {\bibinfo {title} {Forces driving epithelial
  wound healing},\ }\href {https://doi.org/10.1038/nphys3040} {\bibfield
  {journal} {\bibinfo  {journal} {Nature Physics}\ }\textbf {\bibinfo {volume}
  {10}},\ \bibinfo {pages} {683} (\bibinfo {year} {2014})}\BibitemShut
  {NoStop}%
\bibitem [{\citenamefont {Jülicher}\ and\ \citenamefont
  {Eaton}(2017)}]{julicherEmergenceTissueShape2017}%
  \BibitemOpen
  \bibfield  {author} {\bibinfo {author} {\bibfnamefont {F.}~\bibnamefont
  {Jülicher}}\ and\ \bibinfo {author} {\bibfnamefont {S.}~\bibnamefont
  {Eaton}},\ }\bibfield  {title} {\bibinfo {title} {Emergence of tissue shape
  changes from collective cell behaviours},\ }\href
  {https://doi.org/10.1016/j.semcdb.2017.04.004} {\bibfield  {journal}
  {\bibinfo  {journal} {Seminars in Cell \& Developmental Biology}\ }\textbf
  {\bibinfo {volume} {67}},\ \bibinfo {pages} {103} (\bibinfo {year}
  {2017})}\BibitemShut {NoStop}%
\bibitem [{\citenamefont {Salbreux}\ \emph {et~al.}(2012)\citenamefont
  {Salbreux}, \citenamefont {Charras},\ and\ \citenamefont
  {Paluch}}]{salbreuxActinCortexMechanics2012}%
  \BibitemOpen
  \bibfield  {author} {\bibinfo {author} {\bibfnamefont {G.}~\bibnamefont
  {Salbreux}}, \bibinfo {author} {\bibfnamefont {G.}~\bibnamefont {Charras}},\
  and\ \bibinfo {author} {\bibfnamefont {E.}~\bibnamefont {Paluch}},\
  }\bibfield  {title} {\bibinfo {title} {Actin cortex mechanics and cellular
  morphogenesis},\ }\href {https://doi.org/10.1016/j.tcb.2012.07.001}
  {\bibfield  {journal} {\bibinfo  {journal} {Trends in Cell Biology}\ }\textbf
  {\bibinfo {volume} {22}},\ \bibinfo {pages} {536} (\bibinfo {year}
  {2012})}\BibitemShut {NoStop}%
\bibitem [{\citenamefont {Khalilgharibi}\ \emph {et~al.}(2019)\citenamefont
  {Khalilgharibi}, \citenamefont {Fouchard}, \citenamefont {Asadipour},
  \citenamefont {Barrientos}, \citenamefont {Duda}, \citenamefont {Bonfanti},
  \citenamefont {Yonis}, \citenamefont {Harris}, \citenamefont {Mosaffa},
  \citenamefont {Fujita}, \citenamefont {Kabla}, \citenamefont {Mao},
  \citenamefont {Baum}, \citenamefont {Muñoz}, \citenamefont {Miodownik},\
  and\ \citenamefont {Charras}}]{khalilgharibiStressRelaxationEpithelial2019a}%
  \BibitemOpen
  \bibfield  {author} {\bibinfo {author} {\bibfnamefont {N.}~\bibnamefont
  {Khalilgharibi}}, \bibinfo {author} {\bibfnamefont {J.}~\bibnamefont
  {Fouchard}}, \bibinfo {author} {\bibfnamefont {N.}~\bibnamefont {Asadipour}},
  \bibinfo {author} {\bibfnamefont {R.}~\bibnamefont {Barrientos}}, \bibinfo
  {author} {\bibfnamefont {M.}~\bibnamefont {Duda}}, \bibinfo {author}
  {\bibfnamefont {A.}~\bibnamefont {Bonfanti}}, \bibinfo {author}
  {\bibfnamefont {A.}~\bibnamefont {Yonis}}, \bibinfo {author} {\bibfnamefont
  {A.}~\bibnamefont {Harris}}, \bibinfo {author} {\bibfnamefont
  {P.}~\bibnamefont {Mosaffa}}, \bibinfo {author} {\bibfnamefont
  {Y.}~\bibnamefont {Fujita}}, \bibinfo {author} {\bibfnamefont
  {A.}~\bibnamefont {Kabla}}, \bibinfo {author} {\bibfnamefont
  {Y.}~\bibnamefont {Mao}}, \bibinfo {author} {\bibfnamefont {B.}~\bibnamefont
  {Baum}}, \bibinfo {author} {\bibfnamefont {J.~J.}\ \bibnamefont {Muñoz}},
  \bibinfo {author} {\bibfnamefont {M.}~\bibnamefont {Miodownik}},\ and\
  \bibinfo {author} {\bibfnamefont {G.}~\bibnamefont {Charras}},\ }\bibfield
  {title} {\bibinfo {title} {Stress relaxation in epithelial monolayers is
  controlled by the actomyosin cortex},\ }\href
  {https://doi.org/10.1038/s41567-019-0516-6} {\bibfield  {journal} {\bibinfo
  {journal} {Nature Physics}\ }\textbf {\bibinfo {volume} {15}},\ \bibinfo
  {pages} {839} (\bibinfo {year} {2019})}\BibitemShut {NoStop}%
\bibitem [{\citenamefont {Duque}\ \emph {et~al.}(2024)\citenamefont {Duque},
  \citenamefont {Bonfanti}, \citenamefont {Fouchard}, \citenamefont {Baldauf},
  \citenamefont {Azenha}, \citenamefont {Ferber}, \citenamefont {Harris},
  \citenamefont {Barriga}, \citenamefont {Kabla},\ and\ \citenamefont
  {Charras}}]{duqueRuptureStrengthLiving2024}%
  \BibitemOpen
  \bibfield  {author} {\bibinfo {author} {\bibfnamefont {J.}~\bibnamefont
  {Duque}}, \bibinfo {author} {\bibfnamefont {A.}~\bibnamefont {Bonfanti}},
  \bibinfo {author} {\bibfnamefont {J.}~\bibnamefont {Fouchard}}, \bibinfo
  {author} {\bibfnamefont {L.}~\bibnamefont {Baldauf}}, \bibinfo {author}
  {\bibfnamefont {S.~R.}\ \bibnamefont {Azenha}}, \bibinfo {author}
  {\bibfnamefont {E.}~\bibnamefont {Ferber}}, \bibinfo {author} {\bibfnamefont
  {A.}~\bibnamefont {Harris}}, \bibinfo {author} {\bibfnamefont {E.~H.}\
  \bibnamefont {Barriga}}, \bibinfo {author} {\bibfnamefont {A.~J.}\
  \bibnamefont {Kabla}},\ and\ \bibinfo {author} {\bibfnamefont
  {G.}~\bibnamefont {Charras}},\ }\bibfield  {title} {\bibinfo {title} {Rupture
  strength of living cell monolayers},\ }\href
  {https://doi.org/10.1038/s41563-024-02027-3} {\bibfield  {journal} {\bibinfo
  {journal} {Nature Materials}\ }\textbf {\bibinfo {volume} {23}},\ \bibinfo
  {pages} {1563} (\bibinfo {year} {2024})}\BibitemShut {NoStop}%
\bibitem [{\citenamefont {Kong}\ \emph {et~al.}(2019)\citenamefont {Kong},
  \citenamefont {Loison}, \citenamefont {Chavadimane~Shivakumar}, \citenamefont
  {Chan}, \citenamefont {Saadaoui}, \citenamefont {Collinet}, \citenamefont
  {Lenne},\ and\ \citenamefont
  {Clément}}]{kongExperimentalValidationForce2019a}%
  \BibitemOpen
  \bibfield  {author} {\bibinfo {author} {\bibfnamefont {W.}~\bibnamefont
  {Kong}}, \bibinfo {author} {\bibfnamefont {O.}~\bibnamefont {Loison}},
  \bibinfo {author} {\bibfnamefont {P.}~\bibnamefont {Chavadimane~Shivakumar}},
  \bibinfo {author} {\bibfnamefont {E.~H.}\ \bibnamefont {Chan}}, \bibinfo
  {author} {\bibfnamefont {M.}~\bibnamefont {Saadaoui}}, \bibinfo {author}
  {\bibfnamefont {C.}~\bibnamefont {Collinet}}, \bibinfo {author}
  {\bibfnamefont {P.-F.}\ \bibnamefont {Lenne}},\ and\ \bibinfo {author}
  {\bibfnamefont {R.}~\bibnamefont {Clément}},\ }\bibfield  {title} {\bibinfo
  {title} {Experimental validation of force inference in epithelia from cell to
  tissue scale},\ }\href {https://doi.org/10.1038/s41598-019-50690-3}
  {\bibfield  {journal} {\bibinfo  {journal} {Scientific Reports}\ }\textbf
  {\bibinfo {volume} {9}},\ \bibinfo {pages} {14647} (\bibinfo {year}
  {2019})}\BibitemShut {NoStop}%
\bibitem [{\citenamefont {Brauns}\ \emph {et~al.}(2024)\citenamefont {Brauns},
  \citenamefont {Claussen}, \citenamefont {Lefebvre}, \citenamefont
  {Wieschaus},\ and\ \citenamefont
  {Shraiman}}]{braunsGeometricBasisEpithelial2024}%
  \BibitemOpen
  \bibfield  {author} {\bibinfo {author} {\bibfnamefont {F.}~\bibnamefont
  {Brauns}}, \bibinfo {author} {\bibfnamefont {N.~H.}\ \bibnamefont
  {Claussen}}, \bibinfo {author} {\bibfnamefont {M.~F.}\ \bibnamefont
  {Lefebvre}}, \bibinfo {author} {\bibfnamefont {E.~F.}\ \bibnamefont
  {Wieschaus}},\ and\ \bibinfo {author} {\bibfnamefont {B.~I.}\ \bibnamefont
  {Shraiman}},\ }\bibfield  {title} {\bibinfo {title} {The {{Geometric Basis}}
  of {{Epithelial Convergent Extension}}},\ }\bibfield  {journal} {\bibinfo
  {journal} {eLife}\ }\textbf {\bibinfo {volume} {13}},\ \href
  {https://doi.org/10.7554/eLife.95521.2} {10.7554/eLife.95521.2} (\bibinfo
  {year} {2024})\BibitemShut {NoStop}%
\bibitem [{\citenamefont {Kim}\ and\ \citenamefont
  {Hilgenfeldt}(2015)}]{kimCellShapesPatterns2015a}%
  \BibitemOpen
  \bibfield  {author} {\bibinfo {author} {\bibfnamefont {S.}~\bibnamefont
  {Kim}}\ and\ \bibinfo {author} {\bibfnamefont {S.}~\bibnamefont
  {Hilgenfeldt}},\ }\bibfield  {title} {\bibinfo {title} {Cell shapes and
  patterns as quantitative indicators of tissue stress in the plant
  epidermis},\ }\href {https://doi.org/10.1039/C5SM01563D} {\bibfield
  {journal} {\bibinfo  {journal} {Soft Matter}\ }\textbf {\bibinfo {volume}
  {11}},\ \bibinfo {pages} {7270} (\bibinfo {year} {2015})}\BibitemShut
  {NoStop}%
\bibitem [{\citenamefont {Guck}\ \emph {et~al.}(2001)\citenamefont {Guck},
  \citenamefont {Ananthakrishnan}, \citenamefont {Mahmood}, \citenamefont
  {Moon}, \citenamefont {Cunningham},\ and\ \citenamefont
  {Käs}}]{guckOpticalStretcherNovel2001}%
  \BibitemOpen
  \bibfield  {author} {\bibinfo {author} {\bibfnamefont {J.}~\bibnamefont
  {Guck}}, \bibinfo {author} {\bibfnamefont {R.}~\bibnamefont
  {Ananthakrishnan}}, \bibinfo {author} {\bibfnamefont {H.}~\bibnamefont
  {Mahmood}}, \bibinfo {author} {\bibfnamefont {T.~J.}\ \bibnamefont {Moon}},
  \bibinfo {author} {\bibfnamefont {C.~C.}\ \bibnamefont {Cunningham}},\ and\
  \bibinfo {author} {\bibfnamefont {J.}~\bibnamefont {Käs}},\ }\bibfield
  {title} {\bibinfo {title} {The {{Optical Stretcher}}: {{A Novel Laser Tool}}
  to {{Micromanipulate Cells}}},\ }\href
  {https://doi.org/10.1016/S0006-3495(01)75740-2} {\bibfield  {journal}
  {\bibinfo  {journal} {Biophysical Journal}\ }\textbf {\bibinfo {volume}
  {81}},\ \bibinfo {pages} {767} (\bibinfo {year} {2001})}\BibitemShut
  {NoStop}%
\bibitem [{\citenamefont {Serwane}\ \emph {et~al.}(2017)\citenamefont
  {Serwane}, \citenamefont {Mongera}, \citenamefont {Rowghanian}, \citenamefont
  {Kealhofer}, \citenamefont {Lucio}, \citenamefont {Hockenbery},\ and\
  \citenamefont {Campàs}}]{serwaneVivoQuantificationSpatially2017}%
  \BibitemOpen
  \bibfield  {author} {\bibinfo {author} {\bibfnamefont {F.}~\bibnamefont
  {Serwane}}, \bibinfo {author} {\bibfnamefont {A.}~\bibnamefont {Mongera}},
  \bibinfo {author} {\bibfnamefont {P.}~\bibnamefont {Rowghanian}}, \bibinfo
  {author} {\bibfnamefont {D.~A.}\ \bibnamefont {Kealhofer}}, \bibinfo {author}
  {\bibfnamefont {A.~A.}\ \bibnamefont {Lucio}}, \bibinfo {author}
  {\bibfnamefont {Z.~M.}\ \bibnamefont {Hockenbery}},\ and\ \bibinfo {author}
  {\bibfnamefont {O.}~\bibnamefont {Campàs}},\ }\bibfield  {title} {\bibinfo
  {title} {In vivo quantification of spatially varying mechanical properties in
  developing tissues},\ }\href {https://doi.org/10.1038/nmeth.4101} {\bibfield
  {journal} {\bibinfo  {journal} {Nature Methods}\ }\textbf {\bibinfo {volume}
  {14}},\ \bibinfo {pages} {181} (\bibinfo {year} {2017})}\BibitemShut
  {NoStop}%
\bibitem [{\citenamefont {Atia}\ \emph {et~al.}(2018)\citenamefont {Atia},
  \citenamefont {Bi}, \citenamefont {Sharma}, \citenamefont {Mitchel},
  \citenamefont {Gweon}, \citenamefont {A.~Koehler}, \citenamefont {DeCamp},
  \citenamefont {Lan}, \citenamefont {Kim}, \citenamefont {Hirsch},
  \citenamefont {Pegoraro}, \citenamefont {Lee}, \citenamefont {Starr},
  \citenamefont {Weitz}, \citenamefont {Martin}, \citenamefont {Park},
  \citenamefont {Butler},\ and\ \citenamefont
  {Fredberg}}]{atiaGeometricConstraintsEpithelial2018}%
  \BibitemOpen
  \bibfield  {author} {\bibinfo {author} {\bibfnamefont {L.}~\bibnamefont
  {Atia}}, \bibinfo {author} {\bibfnamefont {D.}~\bibnamefont {Bi}}, \bibinfo
  {author} {\bibfnamefont {Y.}~\bibnamefont {Sharma}}, \bibinfo {author}
  {\bibfnamefont {J.~A.}\ \bibnamefont {Mitchel}}, \bibinfo {author}
  {\bibfnamefont {B.}~\bibnamefont {Gweon}}, \bibinfo {author} {\bibfnamefont
  {S.}~\bibnamefont {A.~Koehler}}, \bibinfo {author} {\bibfnamefont {S.~J.}\
  \bibnamefont {DeCamp}}, \bibinfo {author} {\bibfnamefont {B.}~\bibnamefont
  {Lan}}, \bibinfo {author} {\bibfnamefont {J.~H.}\ \bibnamefont {Kim}},
  \bibinfo {author} {\bibfnamefont {R.}~\bibnamefont {Hirsch}}, \bibinfo
  {author} {\bibfnamefont {A.~F.}\ \bibnamefont {Pegoraro}}, \bibinfo {author}
  {\bibfnamefont {K.~H.}\ \bibnamefont {Lee}}, \bibinfo {author} {\bibfnamefont
  {J.~R.}\ \bibnamefont {Starr}}, \bibinfo {author} {\bibfnamefont {D.~A.}\
  \bibnamefont {Weitz}}, \bibinfo {author} {\bibfnamefont {A.~C.}\ \bibnamefont
  {Martin}}, \bibinfo {author} {\bibfnamefont {J.-A.}\ \bibnamefont {Park}},
  \bibinfo {author} {\bibfnamefont {J.~P.}\ \bibnamefont {Butler}},\ and\
  \bibinfo {author} {\bibfnamefont {J.~J.}\ \bibnamefont {Fredberg}},\
  }\bibfield  {title} {\bibinfo {title} {Geometric constraints during
  epithelial jamming},\ }\href {https://doi.org/10.1038/s41567-018-0089-9}
  {\bibfield  {journal} {\bibinfo  {journal} {Nature Physics}\ }\textbf
  {\bibinfo {volume} {14}},\ \bibinfo {pages} {613} (\bibinfo {year}
  {2018})}\BibitemShut {NoStop}%
\bibitem [{\citenamefont {Sánchez‐Gutiérrez}\ \emph
  {et~al.}(2016)\citenamefont {Sánchez‐Gutiérrez}, \citenamefont
  {Tozluoglu}, \citenamefont {Barry}, \citenamefont {Pascual}, \citenamefont
  {Mao},\ and\ \citenamefont
  {Escudero}}]{sanchez-gutierrezFundamentalPhysicalCellular2016}%
  \BibitemOpen
  \bibfield  {author} {\bibinfo {author} {\bibfnamefont {D.}~\bibnamefont
  {Sánchez‐Gutiérrez}}, \bibinfo {author} {\bibfnamefont {M.}~\bibnamefont
  {Tozluoglu}}, \bibinfo {author} {\bibfnamefont {J.~D.}\ \bibnamefont
  {Barry}}, \bibinfo {author} {\bibfnamefont {A.}~\bibnamefont {Pascual}},
  \bibinfo {author} {\bibfnamefont {Y.}~\bibnamefont {Mao}},\ and\ \bibinfo
  {author} {\bibfnamefont {L.~M.}\ \bibnamefont {Escudero}},\ }\bibfield
  {title} {\bibinfo {title} {Fundamental physical cellular constraints drive
  self‐organization of tissues},\ }\href
  {https://doi.org/10.15252/embj.201592374} {\bibfield  {journal} {\bibinfo
  {journal} {The EMBO Journal}\ }\textbf {\bibinfo {volume} {35}},\ \bibinfo
  {pages} {77} (\bibinfo {year} {2016})}\BibitemShut {NoStop}%
\bibitem [{\citenamefont {Sadhukhan}\ and\ \citenamefont
  {Nandi}(2022)}]{sadhukhanOriginUniversalCell2022a}%
  \BibitemOpen
  \bibfield  {author} {\bibinfo {author} {\bibfnamefont {S.}~\bibnamefont
  {Sadhukhan}}\ and\ \bibinfo {author} {\bibfnamefont {S.~K.}\ \bibnamefont
  {Nandi}},\ }\bibfield  {title} {\bibinfo {title} {On the origin of universal
  cell shape variability in confluent epithelial monolayers},\ }\href
  {https://doi.org/10.7554/eLife.76406} {\bibfield  {journal} {\bibinfo
  {journal} {eLife}\ }\textbf {\bibinfo {volume} {11}},\ \bibinfo {pages}
  {e76406} (\bibinfo {year} {2022})}\BibitemShut {NoStop}%
\bibitem [{\citenamefont {Li}\ \emph {et~al.}(2021)\citenamefont {Li},
  \citenamefont {Ibar}, \citenamefont {Zhou}, \citenamefont {Moazzeni},
  \citenamefont {Norris}, \citenamefont {Irvine}, \citenamefont {Liu},\ and\
  \citenamefont {Lin}}]{li2GammaDistributions2021}%
  \BibitemOpen
  \bibfield  {author} {\bibinfo {author} {\bibfnamefont {R.}~\bibnamefont
  {Li}}, \bibinfo {author} {\bibfnamefont {C.}~\bibnamefont {Ibar}}, \bibinfo
  {author} {\bibfnamefont {Z.}~\bibnamefont {Zhou}}, \bibinfo {author}
  {\bibfnamefont {S.}~\bibnamefont {Moazzeni}}, \bibinfo {author}
  {\bibfnamefont {A.~N.}\ \bibnamefont {Norris}}, \bibinfo {author}
  {\bibfnamefont {K.~D.}\ \bibnamefont {Irvine}}, \bibinfo {author}
  {\bibfnamefont {L.}~\bibnamefont {Liu}},\ and\ \bibinfo {author}
  {\bibfnamefont {H.}~\bibnamefont {Lin}},\ }\bibfield  {title} {\bibinfo
  {title} {E 2 and gamma distributions in polygonal networks},\ }\href
  {https://doi.org/10.1103/PhysRevResearch.3.L042001} {\bibfield  {journal}
  {\bibinfo  {journal} {Phys. Rev. Research}\ }\textbf {\bibinfo {volume}
  {3}},\ \bibinfo {pages} {L042001} (\bibinfo {year} {2021})}\BibitemShut
  {NoStop}%
\bibitem [{\citenamefont {Nagai}\ and\ \citenamefont
  {Honda}(2001)}]{nagaiDynamicCellModel2001}%
  \BibitemOpen
  \bibfield  {author} {\bibinfo {author} {\bibfnamefont {T.}~\bibnamefont
  {Nagai}}\ and\ \bibinfo {author} {\bibfnamefont {H.}~\bibnamefont {Honda}},\
  }\bibfield  {title} {\bibinfo {title} {A dynamic cell model for the formation
  of epithelial tissues},\ }\href {https://doi.org/10.1080/13642810108205772}
  {\bibfield  {journal} {\bibinfo  {journal} {Philosophical Magazine B}\
  }\textbf {\bibinfo {volume} {81}},\ \bibinfo {pages} {699} (\bibinfo {year}
  {2001})}\BibitemShut {NoStop}%
\bibitem [{\citenamefont {Farhadifar}\ \emph {et~al.}(2007)\citenamefont
  {Farhadifar}, \citenamefont {Röper}, \citenamefont {Aigouy}, \citenamefont
  {Eaton},\ and\ \citenamefont
  {Jülicher}}]{farhadifarInfluenceCellMechanics2007}%
  \BibitemOpen
  \bibfield  {author} {\bibinfo {author} {\bibfnamefont {R.}~\bibnamefont
  {Farhadifar}}, \bibinfo {author} {\bibfnamefont {J.-C.}\ \bibnamefont
  {Röper}}, \bibinfo {author} {\bibfnamefont {B.}~\bibnamefont {Aigouy}},
  \bibinfo {author} {\bibfnamefont {S.}~\bibnamefont {Eaton}},\ and\ \bibinfo
  {author} {\bibfnamefont {F.}~\bibnamefont {Jülicher}},\ }\bibfield  {title}
  {\bibinfo {title} {The {{Influence}} of {{Cell Mechanics}}, {{Cell-Cell
  Interactions}}, and {{Proliferation}} on {{Epithelial Packing}}},\ }\href
  {https://doi.org/10.1016/j.cub.2007.11.049} {\bibfield  {journal} {\bibinfo
  {journal} {Current Biology}\ }\textbf {\bibinfo {volume} {17}},\ \bibinfo
  {pages} {2095} (\bibinfo {year} {2007})}\BibitemShut {NoStop}%
\bibitem [{\citenamefont {Bi}\ \emph {et~al.}(2015)\citenamefont {Bi},
  \citenamefont {Lopez}, \citenamefont {Schwarz},\ and\ \citenamefont
  {Manning}}]{biDensityindependentRigidityTransition2015}%
  \BibitemOpen
  \bibfield  {author} {\bibinfo {author} {\bibfnamefont {D.}~\bibnamefont
  {Bi}}, \bibinfo {author} {\bibfnamefont {J.~H.}\ \bibnamefont {Lopez}},
  \bibinfo {author} {\bibfnamefont {J.~M.}\ \bibnamefont {Schwarz}},\ and\
  \bibinfo {author} {\bibfnamefont {M.~L.}\ \bibnamefont {Manning}},\
  }\bibfield  {title} {\bibinfo {title} {A density-independent rigidity
  transition in biological tissues},\ }\href
  {https://doi.org/10.1038/nphys3471} {\bibfield  {journal} {\bibinfo
  {journal} {Nature Phys}\ }\textbf {\bibinfo {volume} {11}},\ \bibinfo {pages}
  {1074} (\bibinfo {year} {2015})}\BibitemShut {NoStop}%
\bibitem [{\citenamefont {Park}\ \emph {et~al.}(2015)\citenamefont {Park},
  \citenamefont {Kim}, \citenamefont {Bi}, \citenamefont {Mitchel},
  \citenamefont {Qazvini}, \citenamefont {Tantisira}, \citenamefont {Park},
  \citenamefont {McGill}, \citenamefont {Kim}, \citenamefont {Gweon},
  \citenamefont {Notbohm}, \citenamefont {Steward~Jr}, \citenamefont {Burger},
  \citenamefont {Randell}, \citenamefont {Kho}, \citenamefont {Tambe},
  \citenamefont {Hardin}, \citenamefont {Shore}, \citenamefont {Israel},
  \citenamefont {Weitz}, \citenamefont {Tschumperlin}, \citenamefont {Henske},
  \citenamefont {Weiss}, \citenamefont {Manning}, \citenamefont {Butler},
  \citenamefont {Drazen},\ and\ \citenamefont
  {Fredberg}}]{parkUnjammingCellShape2015}%
  \BibitemOpen
  \bibfield  {author} {\bibinfo {author} {\bibfnamefont {J.-A.}\ \bibnamefont
  {Park}}, \bibinfo {author} {\bibfnamefont {J.~H.}\ \bibnamefont {Kim}},
  \bibinfo {author} {\bibfnamefont {D.}~\bibnamefont {Bi}}, \bibinfo {author}
  {\bibfnamefont {J.~A.}\ \bibnamefont {Mitchel}}, \bibinfo {author}
  {\bibfnamefont {N.~T.}\ \bibnamefont {Qazvini}}, \bibinfo {author}
  {\bibfnamefont {K.}~\bibnamefont {Tantisira}}, \bibinfo {author}
  {\bibfnamefont {C.~Y.}\ \bibnamefont {Park}}, \bibinfo {author}
  {\bibfnamefont {M.}~\bibnamefont {McGill}}, \bibinfo {author} {\bibfnamefont
  {S.-H.}\ \bibnamefont {Kim}}, \bibinfo {author} {\bibfnamefont
  {B.}~\bibnamefont {Gweon}}, \bibinfo {author} {\bibfnamefont
  {J.}~\bibnamefont {Notbohm}}, \bibinfo {author} {\bibfnamefont
  {R.}~\bibnamefont {Steward~Jr}}, \bibinfo {author} {\bibfnamefont
  {S.}~\bibnamefont {Burger}}, \bibinfo {author} {\bibfnamefont {S.~H.}\
  \bibnamefont {Randell}}, \bibinfo {author} {\bibfnamefont {A.~T.}\
  \bibnamefont {Kho}}, \bibinfo {author} {\bibfnamefont {D.~T.}\ \bibnamefont
  {Tambe}}, \bibinfo {author} {\bibfnamefont {C.}~\bibnamefont {Hardin}},
  \bibinfo {author} {\bibfnamefont {S.~A.}\ \bibnamefont {Shore}}, \bibinfo
  {author} {\bibfnamefont {E.}~\bibnamefont {Israel}}, \bibinfo {author}
  {\bibfnamefont {D.~A.}\ \bibnamefont {Weitz}}, \bibinfo {author}
  {\bibfnamefont {D.~J.}\ \bibnamefont {Tschumperlin}}, \bibinfo {author}
  {\bibfnamefont {E.~P.}\ \bibnamefont {Henske}}, \bibinfo {author}
  {\bibfnamefont {S.~T.}\ \bibnamefont {Weiss}}, \bibinfo {author}
  {\bibfnamefont {M.~L.}\ \bibnamefont {Manning}}, \bibinfo {author}
  {\bibfnamefont {J.~P.}\ \bibnamefont {Butler}}, \bibinfo {author}
  {\bibfnamefont {J.~M.}\ \bibnamefont {Drazen}},\ and\ \bibinfo {author}
  {\bibfnamefont {J.~J.}\ \bibnamefont {Fredberg}},\ }\bibfield  {title}
  {\bibinfo {title} {Unjamming and cell shape in the asthmatic airway
  epithelium},\ }\href {https://doi.org/10.1038/nmat4357} {\bibfield  {journal}
  {\bibinfo  {journal} {Nature Materials}\ }\textbf {\bibinfo {volume} {14}},\
  \bibinfo {pages} {1040} (\bibinfo {year} {2015})}\BibitemShut {NoStop}%
\bibitem [{\citenamefont {Voronoi}(1908)}]{Voronoi1908}%
  \BibitemOpen
  \bibfield  {author} {\bibinfo {author} {\bibfnamefont {G.}~\bibnamefont
  {Voronoi}},\ }\bibfield  {title} {\bibinfo {title} {Nouvelles applications
  des paramètres continus à la théorie des formes quadratiques. deuxième
  mémoire. recherches sur les parallélloèdres primitifs.},\ }\href@noop {}
  {\bibfield  {journal} {\bibinfo  {journal} {Journal für die reine und
  angewandte Mathematik}\ }\textbf {\bibinfo {volume} {134}},\ \bibinfo {pages}
  {198} (\bibinfo {year} {1908})}\BibitemShut {NoStop}%
\bibitem [{\citenamefont {Honda}(1978)}]{hondaDescriptionCellularPatterns1978}%
  \BibitemOpen
  \bibfield  {author} {\bibinfo {author} {\bibfnamefont {H.}~\bibnamefont
  {Honda}},\ }\bibfield  {title} {\bibinfo {title} {Description of cellular
  patterns by {{Dirichlet}} domains: {{The}} two-dimensional case},\ }\href
  {https://doi.org/10.1016/0022-5193(78)90315-6} {\bibfield  {journal}
  {\bibinfo  {journal} {Journal of Theoretical Biology}\ }\textbf {\bibinfo
  {volume} {72}},\ \bibinfo {pages} {523} (\bibinfo {year} {1978})}\BibitemShut
  {NoStop}%
\bibitem [{\citenamefont {Gudipaty}\ \emph {et~al.}(2017)\citenamefont
  {Gudipaty}, \citenamefont {Lindblom}, \citenamefont {Loftus}, \citenamefont
  {Redd}, \citenamefont {Edes}, \citenamefont {Davey}, \citenamefont
  {Krishnegowda},\ and\ \citenamefont
  {Rosenblatt}}]{gudipatyMechanicalStretchTriggers2017}%
  \BibitemOpen
  \bibfield  {author} {\bibinfo {author} {\bibfnamefont {S.~A.}\ \bibnamefont
  {Gudipaty}}, \bibinfo {author} {\bibfnamefont {J.}~\bibnamefont {Lindblom}},
  \bibinfo {author} {\bibfnamefont {P.~D.}\ \bibnamefont {Loftus}}, \bibinfo
  {author} {\bibfnamefont {M.~J.}\ \bibnamefont {Redd}}, \bibinfo {author}
  {\bibfnamefont {K.}~\bibnamefont {Edes}}, \bibinfo {author} {\bibfnamefont
  {C.~F.}\ \bibnamefont {Davey}}, \bibinfo {author} {\bibfnamefont
  {V.}~\bibnamefont {Krishnegowda}},\ and\ \bibinfo {author} {\bibfnamefont
  {J.}~\bibnamefont {Rosenblatt}},\ }\bibfield  {title} {\bibinfo {title}
  {Mechanical stretch triggers rapid epithelial cell division through
  {{Piezo1}}},\ }\href {https://doi.org/10.1038/nature21407} {\bibfield
  {journal} {\bibinfo  {journal} {Nature}\ }\textbf {\bibinfo {volume} {543}},\
  \bibinfo {pages} {118} (\bibinfo {year} {2017})}\BibitemShut {NoStop}%
\bibitem [{\citenamefont {Rappel}\ \emph {et~al.}(1999)\citenamefont {Rappel},
  \citenamefont {Fenton},\ and\ \citenamefont
  {Karma}}]{rappelSpatiotemporalControlWave1999}%
  \BibitemOpen
  \bibfield  {author} {\bibinfo {author} {\bibfnamefont {W.-J.}\ \bibnamefont
  {Rappel}}, \bibinfo {author} {\bibfnamefont {F.}~\bibnamefont {Fenton}},\
  and\ \bibinfo {author} {\bibfnamefont {A.}~\bibnamefont {Karma}},\ }\bibfield
   {title} {\bibinfo {title} {Spatiotemporal {{Control}} of {{Wave
  Instabilities}} in {{Cardiac Tissue}}},\ }\href
  {https://doi.org/10.1103/PhysRevLett.83.456} {\bibfield  {journal} {\bibinfo
  {journal} {Physical Review Letters}\ }\textbf {\bibinfo {volume} {83}},\
  \bibinfo {pages} {456} (\bibinfo {year} {1999})}\BibitemShut {NoStop}%
\bibitem [{\citenamefont {Killeen}\ \emph {et~al.}(2021)\citenamefont
  {Killeen}, \citenamefont {Bertrand},\ and\ \citenamefont
  {Lee}}]{killeenPolarFluctuationsLead2021}%
  \BibitemOpen
  \bibfield  {author} {\bibinfo {author} {\bibfnamefont {A.}~\bibnamefont
  {Killeen}}, \bibinfo {author} {\bibfnamefont {T.}~\bibnamefont {Bertrand}},\
  and\ \bibinfo {author} {\bibfnamefont {C.~F.}\ \bibnamefont {Lee}},\ }\href
  {https://doi.org/10.48550/arXiv.2107.03838} {\bibinfo {title} {Polar
  {{Fluctuations Lead}} to {{Extensile Nematic Behavior}} in {{Confluent
  Tissues}}}} (\bibinfo {year} {2021}),\ \Eprint
  {https://arxiv.org/abs/2107.03838} {2107.03838} \BibitemShut {NoStop}%
\bibitem [{\citenamefont {Srinivasan}\ \emph {et~al.}(2024)\citenamefont
  {Srinivasan}, \citenamefont {Höhn},\ and\ \citenamefont
  {Goldstein}}]{srinivasanStochasticVoronoiTessellations2024}%
  \BibitemOpen
  \bibfield  {author} {\bibinfo {author} {\bibfnamefont {A.}~\bibnamefont
  {Srinivasan}}, \bibinfo {author} {\bibfnamefont {S.~S. M.~H.}\ \bibnamefont
  {Höhn}},\ and\ \bibinfo {author} {\bibfnamefont {R.~E.}\ \bibnamefont
  {Goldstein}},\ }\href {https://doi.org/10.1101/2024.03.11.584390} {\bibinfo
  {title} {Stochastic {{Voronoi Tessellations}} as {{Models}} for {{Cellular
  Neighborhoods}} in {{Simple Multicellular Organisms}}}} (\bibinfo {year}
  {2024})\BibitemShut {NoStop}%
\bibitem [{\citenamefont {Huang}\ \emph {et~al.}(2022)\citenamefont {Huang},
  \citenamefont {Cochran}, \citenamefont {Fielding}, \citenamefont
  {Marchetti},\ and\ \citenamefont
  {Bi}}]{huangShearDrivenSolidificationNonlinear2022}%
  \BibitemOpen
  \bibfield  {author} {\bibinfo {author} {\bibfnamefont {J.}~\bibnamefont
  {Huang}}, \bibinfo {author} {\bibfnamefont {J.~O.}\ \bibnamefont {Cochran}},
  \bibinfo {author} {\bibfnamefont {S.~M.}\ \bibnamefont {Fielding}}, \bibinfo
  {author} {\bibfnamefont {M.~C.}\ \bibnamefont {Marchetti}},\ and\ \bibinfo
  {author} {\bibfnamefont {D.}~\bibnamefont {Bi}},\ }\bibfield  {title}
  {\bibinfo {title} {Shear-{{Driven Solidification}} and {{Nonlinear
  Elasticity}} in {{Epithelial Tissues}}},\ }\href
  {https://doi.org/10.1103/PhysRevLett.128.178001} {\bibfield  {journal}
  {\bibinfo  {journal} {Phys. Rev. Lett.}\ }\textbf {\bibinfo {volume} {128}},\
  \bibinfo {pages} {178001} (\bibinfo {year} {2022})}\BibitemShut {NoStop}%
\bibitem [{Cen()}]{CentroidalVoronoiTessellations}%
  \BibitemOpen
  \href {https://doi.org/10.1137/S0036144599352836} {\bibinfo {title}
  {Centroidal {{Voronoi Tessellations}}: {{Applications}} and
  {{Algorithms}}}}\BibitemShut {NoStop}%
\bibitem [{\citenamefont {Hernandez}\ \emph {et~al.}(2023)\citenamefont
  {Hernandez}, \citenamefont {Staddon}, \citenamefont {Moshe},\ and\
  \citenamefont {Marchetti}}]{hernandezFiniteElasticityVertex2023}%
  \BibitemOpen
  \bibfield  {author} {\bibinfo {author} {\bibfnamefont {A.}~\bibnamefont
  {Hernandez}}, \bibinfo {author} {\bibfnamefont {M.~F.}\ \bibnamefont
  {Staddon}}, \bibinfo {author} {\bibfnamefont {M.}~\bibnamefont {Moshe}},\
  and\ \bibinfo {author} {\bibfnamefont {M.~C.}\ \bibnamefont {Marchetti}},\
  }\bibfield  {title} {\bibinfo {title} {Finite elasticity of the vertex model
  and its role in rigidity of curved cellular tissues},\ }\href
  {https://doi.org/10.1039/D3SM00874F} {\bibfield  {journal} {\bibinfo
  {journal} {Soft Matter}\ }\textbf {\bibinfo {volume} {19}},\ \bibinfo {pages}
  {7744} (\bibinfo {year} {2023})}\BibitemShut {NoStop}%
\bibitem [{\citenamefont {Nestor-Bergmann}\ \emph {et~al.}(2018)\citenamefont
  {Nestor-Bergmann}, \citenamefont {Goddard}, \citenamefont {Woolner},\ and\
  \citenamefont {Jensen}}]{nestor-bergmannRelatingCellShape2018}%
  \BibitemOpen
  \bibfield  {author} {\bibinfo {author} {\bibfnamefont {A.}~\bibnamefont
  {Nestor-Bergmann}}, \bibinfo {author} {\bibfnamefont {G.}~\bibnamefont
  {Goddard}}, \bibinfo {author} {\bibfnamefont {S.}~\bibnamefont {Woolner}},\
  and\ \bibinfo {author} {\bibfnamefont {O.~E.}\ \bibnamefont {Jensen}},\
  }\bibfield  {title} {\bibinfo {title} {Relating cell shape and mechanical
  stress in a spatially disordered epithelium using a vertex-based model},\
  }\href {https://doi.org/10.1093/imammb/dqx008} {\bibfield  {journal}
  {\bibinfo  {journal} {Mathematical Medicine and Biology: A Journal of the
  IMA}\ }\textbf {\bibinfo {volume} {35}},\ \bibinfo {pages} {i1} (\bibinfo
  {year} {2018})}\BibitemShut {NoStop}%
\bibitem [{\citenamefont {De~Luca}\ \emph {et~al.}(2014)\citenamefont
  {De~Luca}, \citenamefont {Zaccaria}, \citenamefont {Hadhoud}, \citenamefont
  {Rizzo}, \citenamefont {Ponzini}, \citenamefont {Morbiducci},\ and\
  \citenamefont {Santoro}}]{delucaZebraBeatFlexiblePlatform2014}%
  \BibitemOpen
  \bibfield  {author} {\bibinfo {author} {\bibfnamefont {E.}~\bibnamefont
  {De~Luca}}, \bibinfo {author} {\bibfnamefont {G.~M.}\ \bibnamefont
  {Zaccaria}}, \bibinfo {author} {\bibfnamefont {M.}~\bibnamefont {Hadhoud}},
  \bibinfo {author} {\bibfnamefont {G.}~\bibnamefont {Rizzo}}, \bibinfo
  {author} {\bibfnamefont {R.}~\bibnamefont {Ponzini}}, \bibinfo {author}
  {\bibfnamefont {U.}~\bibnamefont {Morbiducci}},\ and\ \bibinfo {author}
  {\bibfnamefont {M.~M.}\ \bibnamefont {Santoro}},\ }\bibfield  {title}
  {\bibinfo {title} {{{ZebraBeat}}: A flexible platform for the analysis of the
  cardiac rate in zebrafish embryos},\ }\href
  {https://doi.org/10.1038/srep04898} {\bibfield  {journal} {\bibinfo
  {journal} {Scientific Reports}\ }\textbf {\bibinfo {volume} {4}},\ \bibinfo
  {pages} {4898} (\bibinfo {year} {2014})}\BibitemShut {NoStop}%
\bibitem [{\citenamefont {Erdemci-Tandogan}\ \emph {et~al.}(2018)\citenamefont
  {Erdemci-Tandogan}, \citenamefont {Clark}, \citenamefont {Amack},\ and\
  \citenamefont {Manning}}]{erdemci-tandoganTissueFlowInduces2018b}%
  \BibitemOpen
  \bibfield  {author} {\bibinfo {author} {\bibfnamefont {G.}~\bibnamefont
  {Erdemci-Tandogan}}, \bibinfo {author} {\bibfnamefont {M.~J.}\ \bibnamefont
  {Clark}}, \bibinfo {author} {\bibfnamefont {J.~D.}\ \bibnamefont {Amack}},\
  and\ \bibinfo {author} {\bibfnamefont {M.~L.}\ \bibnamefont {Manning}},\
  }\bibfield  {title} {\bibinfo {title} {Tissue {{Flow Induces Cell Shape
  Changes During Organogenesis}}},\ }\href
  {https://doi.org/10.1016/j.bpj.2018.10.028} {\bibfield  {journal} {\bibinfo
  {journal} {Biophysical Journal}\ }\textbf {\bibinfo {volume} {115}},\
  \bibinfo {pages} {2259} (\bibinfo {year} {2018})}\BibitemShut {NoStop}%
\bibitem [{\citenamefont {Wang}\ \emph {et~al.}(2020)\citenamefont {Wang},
  \citenamefont {Merkel}, \citenamefont {Sutter}, \citenamefont
  {Erdemci-Tandogan}, \citenamefont {Manning},\ and\ \citenamefont
  {Kasza}}]{wangAnisotropyLinksCell2020}%
  \BibitemOpen
  \bibfield  {author} {\bibinfo {author} {\bibfnamefont {X.}~\bibnamefont
  {Wang}}, \bibinfo {author} {\bibfnamefont {M.}~\bibnamefont {Merkel}},
  \bibinfo {author} {\bibfnamefont {L.~B.}\ \bibnamefont {Sutter}}, \bibinfo
  {author} {\bibfnamefont {G.}~\bibnamefont {Erdemci-Tandogan}}, \bibinfo
  {author} {\bibfnamefont {M.~L.}\ \bibnamefont {Manning}},\ and\ \bibinfo
  {author} {\bibfnamefont {K.~E.}\ \bibnamefont {Kasza}},\ }\bibfield  {title}
  {\bibinfo {title} {Anisotropy links cell shapes to tissue flow during
  convergent extension},\ }\href {https://doi.org/10.1073/pnas.1916418117}
  {\bibfield  {journal} {\bibinfo  {journal} {Proceedings of the National
  Academy of Sciences}\ }\textbf {\bibinfo {volume} {117}},\ \bibinfo {pages}
  {13541} (\bibinfo {year} {2020})}\BibitemShut {NoStop}%
\bibitem [{\citenamefont {Petridou}\ \emph {et~al.}(2021)\citenamefont
  {Petridou}, \citenamefont {Corominas-Murtra}, \citenamefont {Heisenberg},\
  and\ \citenamefont {Hannezo}}]{petridouRigidityPercolationUncovers2021}%
  \BibitemOpen
  \bibfield  {author} {\bibinfo {author} {\bibfnamefont {N.~I.}\ \bibnamefont
  {Petridou}}, \bibinfo {author} {\bibfnamefont {B.}~\bibnamefont
  {Corominas-Murtra}}, \bibinfo {author} {\bibfnamefont {C.-P.}\ \bibnamefont
  {Heisenberg}},\ and\ \bibinfo {author} {\bibfnamefont {E.}~\bibnamefont
  {Hannezo}},\ }\bibfield  {title} {\bibinfo {title} {Rigidity percolation
  uncovers a structural basis for embryonic tissue phase transitions},\ }\href
  {https://doi.org/10.1016/j.cell.2021.02.017} {\bibfield  {journal} {\bibinfo
  {journal} {Cell}\ }\textbf {\bibinfo {volume} {184}},\ \bibinfo {pages}
  {1914} (\bibinfo {year} {2021})}\BibitemShut {NoStop}%
\bibitem [{\citenamefont {Mongera}\ \emph {et~al.}(2018)\citenamefont
  {Mongera}, \citenamefont {Rowghanian}, \citenamefont {Gustafson},
  \citenamefont {Shelton}, \citenamefont {Kealhofer}, \citenamefont {Carn},
  \citenamefont {Serwane}, \citenamefont {Lucio}, \citenamefont {Giammona},\
  and\ \citenamefont {Campàs}}]{mongeraFluidtosolidJammingTransition2018}%
  \BibitemOpen
  \bibfield  {author} {\bibinfo {author} {\bibfnamefont {A.}~\bibnamefont
  {Mongera}}, \bibinfo {author} {\bibfnamefont {P.}~\bibnamefont {Rowghanian}},
  \bibinfo {author} {\bibfnamefont {H.~J.}\ \bibnamefont {Gustafson}}, \bibinfo
  {author} {\bibfnamefont {E.}~\bibnamefont {Shelton}}, \bibinfo {author}
  {\bibfnamefont {D.~A.}\ \bibnamefont {Kealhofer}}, \bibinfo {author}
  {\bibfnamefont {E.~K.}\ \bibnamefont {Carn}}, \bibinfo {author}
  {\bibfnamefont {F.}~\bibnamefont {Serwane}}, \bibinfo {author} {\bibfnamefont
  {A.~A.}\ \bibnamefont {Lucio}}, \bibinfo {author} {\bibfnamefont
  {J.}~\bibnamefont {Giammona}},\ and\ \bibinfo {author} {\bibfnamefont
  {O.}~\bibnamefont {Campàs}},\ }\bibfield  {title} {\bibinfo {title} {A
  fluid-to-solid jamming transition underlies vertebrate body axis
  elongation},\ }\href {https://doi.org/10.1038/s41586-018-0479-2} {\bibfield
  {journal} {\bibinfo  {journal} {Nature}\ }\textbf {\bibinfo {volume} {561}},\
  \bibinfo {pages} {401} (\bibinfo {year} {2018})}\BibitemShut {NoStop}%
\bibitem [{\citenamefont {Shellard}\ and\ \citenamefont
  {Mayor}(2023)}]{shellardSculptingStiffnessRigidity2023}%
  \BibitemOpen
  \bibfield  {author} {\bibinfo {author} {\bibfnamefont {A.}~\bibnamefont
  {Shellard}}\ and\ \bibinfo {author} {\bibfnamefont {R.}~\bibnamefont
  {Mayor}},\ }\bibfield  {title} {\bibinfo {title} {Sculpting with stiffness:
  Rigidity as a regulator of morphogenesis},\ }\href
  {https://doi.org/10.1042/BST20220826} {\bibfield  {journal} {\bibinfo
  {journal} {Biochemical Society Transactions}\ }\textbf {\bibinfo {volume}
  {51}},\ \bibinfo {pages} {1009} (\bibinfo {year} {2023})}\BibitemShut
  {NoStop}%
\end{thebibliography}

%

\end{document}


\title{Supplemetary Material}

\maketitle
\section{Vertex Model}

The energy function is a sum of contributions from N cells.

\begin{equation}
    E=\sum^N_{a=1} K_A(A_a-A_0)^2+K_P(P_a-P_0)^2
\end{equation}

The force is calculated from the energy functional as $\mathbf{F}_i = -\nabla_i E$

\begin{align}   
    \mathbf{F}_i &= -\frac{\partial E}{\partial \mathbf{r}_i}
    = -\sum_{a \in\mathcal{N}_i} \frac{\partial E_a}{\partial \mathbf{r}_i}    
\end{align}

where $\mathbf{r}_i$ is the position of vertex {\it i}, and the sum is over cells $a \in \mathcal{N}_i$, the three cells associated with vertex {\it i}

The energy derivatives are
\begin{align}
\frac{\partial E_a}{\partial \mathbf{r}_i} 
&= 
\frac{\partial E_a}{\partial A_a} \frac{\partial A_a}{\partial \mathbf{r}_i}
+ \frac{\partial E_a}{\partial P_a} \frac{\partial P_a}{\partial \mathbf{r}_i} \nonumber \\[6pt]
&= 
2K_A (A_a - A_0) \frac{\partial A_a}{\partial \mathbf{r}_i}
+ 2K_P (P_a - P_0) \frac{\partial P_a}{\partial \mathbf{r}_i},
\label{eq:force_derivative}
\end{align}

The area and perimeter derivatives are given by
\begin{align}
\frac{\partial A_a}{\partial \mathbf{r}_i} &= 
\frac{1}{2}\left( |\mathbf{r}_{ij}|\,\hat{\mathbf{n}}_{ij} + |\mathbf{r}_{ik}|\,\hat{\mathbf{n}}_{ik} \right), \\[6pt]
\frac{\partial P_a}{\partial \mathbf{r}_i} &= 
\hat{\mathbf{r}}_{ij} + \hat{\mathbf{r}}_{ik},
\label{eq:area_perimeter_derivatives}
\end{align}
where vertices \( j \) and \( k \) are the two vertices directly before and after \( i \) (respectively)
when traversing the vertices of cell \( a \) in a clockwise loop. 
We denote \( \mathbf{r}_{ij} = \mathbf{r}_i - \mathbf{r}_j \) as the cell edge connecting vertices \( i \) and \( j \),
\( \hat{\mathbf{n}}_{ij} \) as the outward-facing normal unit vector to that edge,
and \( \hat{\mathbf{r}}_{ij} = \mathbf{r}_{ij} / |\mathbf{r}_{ij}| \).

We use the dynamic implementation of the vertex model, in which the tissue mechanical force on each vertex is balanced by substrate friction (overdamped dynamics)

\begin{align}
\frac{d \mathbf{r}_i}{d t}&= \frac{1}{\zeta}\mathbf{F}_i
\label{eq: dynamic equation}
\end{align}

where $\zeta$ is the damping coefficient.

\subsection{Model implementation}

For all results we simulate $N=256$ cells on a periodic 2D honeycomb lattice for $\sqrt{N} \times \sqrt{N} $ cells which ensures a total area of $N$ and $\langle A \rangle =1$. We set $K_A=K_P=1$ 

We evolve the vertices by numerically integrating \ref{eq: dynamic equation} over discrete timesteps $\Delta t = 0.01$. 

The unit of time $\tau$ is determined by $\zeta$, which we vary between cyclic and static stretch conditions. 

\blankline
We implement stretch isotropically by expanding the coordinates of all vertices at the start of each timestep before calculating forces and evolving vertex positions.

\blankline
To initialise homogeneous cells at a given $p_0$, we adapt the protocol of Killeen et al. Specifically, we include the motility mechanism described in \cite{SkilleenPolarFluctuationsLead2021}, initialising cells with random polarities, and we allow the system to run with motility for $3 \times 10^3$ time steps. We then remove the motility and allow the system to relax for a further $3 \times 10^3$ time steps. For all values of $p_0$ this ensures homogeneous states with $p=p_0$ for all cells.

\blankline

\subsection{Stretch implementation}

We first generate our homogeneous cells at a given $p_0$ as described above, and then begin the stretching simulations.

\blankline

\textbf{Cyclic stretch implementation}

\blankline

For cyclic stretch simulations, we define a stretch factor $e$, and isotropically stretch and relax cells sinusoidally to a maximum expansion of $1+e$ over a time period $T_H$, which is chosen to correspond to 50 time steps, $T_H=50 \Delta t$. Our cyclic stretch simulations are inspired by the zebrafish heartbeat, which has a frequency of $2 \sim 3 \text{ Hz}$. Combined with previous estimates for $\tau \approx 1$ minute \cite{Serdemci-tandoganTissueFlowInduces2018b}, we choose $\zeta = 100$.

\blankline

\textbf{Static stretch implementation}

\blankline

For static stretch simulations, in which cells undergo an isotropic expansion which is maintained, we expand uniformly to $1+e$ over $1 \times 10^3$ timesteps. $\zeta = 1$ for static stretch simulations as there are no further time dependent mechanisms.

\section{CVT measurements}

We use a cell-wise vertex displacement measurement to quantify deviation from the centroidal tessellation, which we call $\tri_{\rm cVT}$.

For our vertex-model networks, we perform Voronoi tessellations about cell centroids and compute the CVT displacement for a cell {\it a} as:

\begin{align}  
    \tri_{{\rm cVT} a}&=\frac{1}{N_a\sqrt{A_a}}\sum_{i=1}^{N_a} \sqrt{(\mathbf{r}_i-\mathbf{r}^v_i)^2} 
\end{align}

where $A_a$ is the area of the cell, $N_a$ is the number of vertices, $\mathbf{r}_i$ refers to a cell vertex and $\mathbf{r}_i^v$ refers to the corresponding voronoi vertex.
In the case where the voronoi cell and VM cell do not have a corresponding vertex (i.e they have swapped junctions) we compute an average junction displacement dependent on the deviation of midpoint of swapped edges.

The measure is then averaged over all cells
\begin{align}
    \tri_{\rm cVT}&=\frac{1}{N}\sum^N_{i=1} \tri_{{\rm cVT} a}   
\end{align}

\section{Analytical results}

In recent work by Huang et al. \cite{ShuangShearDrivenSolidificationNonlinear2022} extending earlier work on single-cell mean-field vertex models \cite{SczajkowskiHydrodynamicsShapedrivenRigidity2018b}, polygons are approximated by an area-preserving deformation tensor, which between general polygon vertices $\mathbf{y}$ and co-centered regular n-gon vertices $\mathbf{x}$ as $\mathbf{y}=\mathbf{Dx}$:
\begin{align*}
    \mathbf{D} &= 
    \begin{pmatrix}
        D_{xx} & 0 \\
        D_{xy} & 1/D_{xx}
    \end{pmatrix}
    &&
    \begin{aligned}
        &D_{xx}-1 = M(\theta)\cos\theta \\
        &D_{xy} = M(\theta)\sin\theta
    \end{aligned}
\end{align*}
The perimeter of a deformed n-gon can be approximated by $\mathbf{D}$ as
\begin{align}
P \approx P_{\rm reg}(n) + \frac{15}{32}P_{\rm reg}(n)[1+\frac{3}{5}\cos{2\theta}]M(\theta)^2
\end{align}

By assuming area-preserving deformations only, the single-cell vertex model energy can be described by the perimeter component, which using (9) takes the form 
\begin{equation}
E_{\text{cell}} = \frac{1}{2}\alpha t\, m(\theta,M)^2 + \frac{1}{4}\beta m(\theta,M)^4,
\end{equation}
where $m(\theta,M)=[1+\frac{3}{5}\cos{2\theta}]^{\frac{1}{2}}$

\blankline

Our goal is to rewrite the single cell mean field energy in terms of $\|\mathbf{D}\|^2$, to relate it to $E_q$, which, approximating polygons by $\mathbf{D}$, takes the form $E_q\propto\sum_{i=1}^N m_0 \|\mathbf{D}_i\|^2$ \cite{Sli2GammaDistributions2021}.

We define $\mathbf{C}=\mathbf{D}-\mathbf{I}$ so that $C_{\rm xx}=D_{\rm xx}-1$, $C_{\rm xy}=D_{\rm xy}$ and $C_{\rm yy}=\frac{1}{D_{\rm xx}-1}$. 

$m^2(\theta,M)$ then takes the simple form $m^2(\theta,M)=4C^2_{\rm xx}+C^2_{\rm xy}$.
\blankline

For area-preserving $\mathbf{D}$, the frobenius norm is given by 
\begin{align}
&\|D\|^2=D_{xx}^2+D_{xy}^2+\frac{1}{D_{xx}^2} \\
&\|D\|^2=(1+C_{xx})^2+C_{xy}^2+(1+C_{xx})^{-2} \\
&\|D\|^2=4C_{xx}^2+C_{xy}^2
\end{align}
to quadratic order in $C_{xx}$, therefore $m^2(\theta,M)=\|D\|^2-2$, which relates to the full strain energy.

$E_{vm}\propto p_0^2(\|D\|^2-2)$ and therefore $E_q=\sum \frac{m_0}{p^2_0}E_{vm}$

Regular n-gons of constant area satisfy $\frac{m_0}{p^2_0}\propto \frac{1}{n^2}[1+3\cot{\frac{\pi}{n}}]$, which for n=5-7 varies as 0.267, 0.278, 0.284 (within 3 percent).

\section{Static stretch and quenched $p_0$ results}
In the main text we present results for how the VM-CVT relationship is dependent predominantly on shifting $p_0^{eff}$ under cyclic stretching conditions. Due to the separation in timescales we expect the behaviour to be equivalent to cells feeling a time-averaged static stretch, which is supported by the good curve collapse of data by the predicted time averaged $\langle p_0^{eff}(t)\rangle$, Fig. 2c. Therefore we expect the qualitative results to be equivalent under both static-stretching and the direct quenching of $p_0$, which we validate in Fig. S3 and Fig. S4. 

In Fig. S3 we show how under static-stretch we see a very good curve collapse of $\tri_{\rm cVT}$ on $p_0^{eff}$ in the fluid phase. For a fractional stretch $e$ we can exactly predict the effective shape index as $p_0^{eff}=\frac{p_0}{1+e}$. 

In Fig. S4 we present the results for quenching $p_0$. We simulate initial homogeneous states at a given initial value of $p_0$, and then steadily reduce $p_0$ to a target value $p^t_0$. We recapitulate the observed features of stretching, specifically that cell layers are hard to distinguish by mean observable quantities $\tri_{\rm cVT}$ and $\langle p \rangle$ (Fig. S4C), and that the measure of structural disorder $P(6)$ can distinguish the extent of underlying quench for a given observed shape index $\langle p \rangle$.

\begin{figure}[t]  
  \centering
  \includegraphics[width=0.7\linewidth]{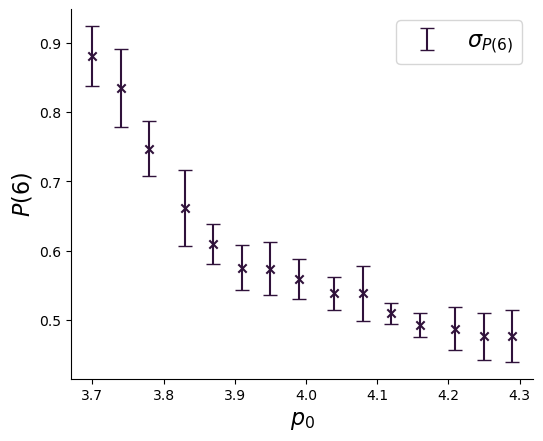}
  \caption{The fraction of hexagons P(6) decreases with increasing $p_0$ for our homogeneous simulations, showing that structural isotropy decreases as $p_0$ increases.}
  \label{SM: S1}
\end{figure}

\begin{figure}[h]  
  \centering
  \includegraphics[width=0.7\linewidth]{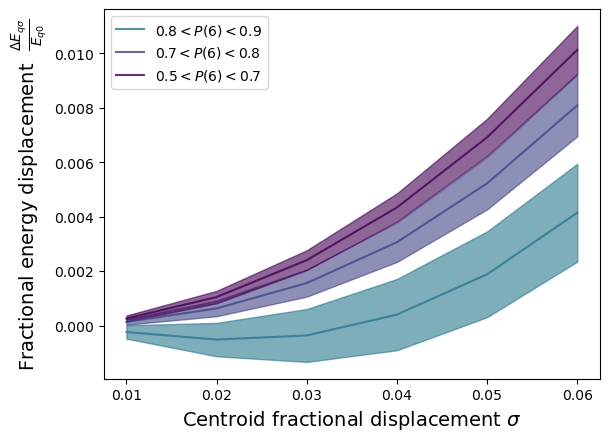}
  \caption{{\it{Disordered solid phase vertex model networks are in steeper local $E_q$ minima.}} For all cyclic stretch simulations in the observable solid phase $\langle p \rangle < 3.81$, we locally perturb centroids by a fractional displacement $\sigma$ and calculate the fractional change in $E_q$. As isotropy decreases, measured by $P(6)$, the local steepness of $E_q$ increases. 
  For each cell-network analysed, we perturb each centroid by a random displacement drawn from a normal distribution with standard deviation $\sigma$. For each network at value of $\sigma$, we obtain a mean value in $\frac{\Delta E_{q \sigma}}{E_{q0}}$ from 100 repeats. Errors represent the standard deviation in the mean of $\frac{\Delta E_{q \sigma}}{E_{q0}}$ for all networks.}
  \label{SM: S2}
\end{figure}

\begin{figure}[t]  
  \centering
  \includegraphics[width=0.5\linewidth]{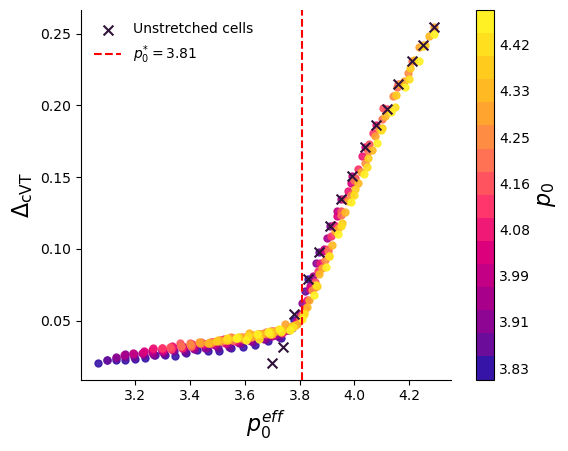}
  \caption{Good data collapse in the fluid regime for all static stretch conditions when plotted in terms of $p_0^{eff}=\frac{p_0}{1+e}$.  $\langle p \rangle$ cannot decrease below 3.72, which corresponds to a honeycomb lattice.}
  \label{SM: S3}
\end{figure}

\begin{figure}[h]  
  \centering
  \includegraphics[width=\linewidth]{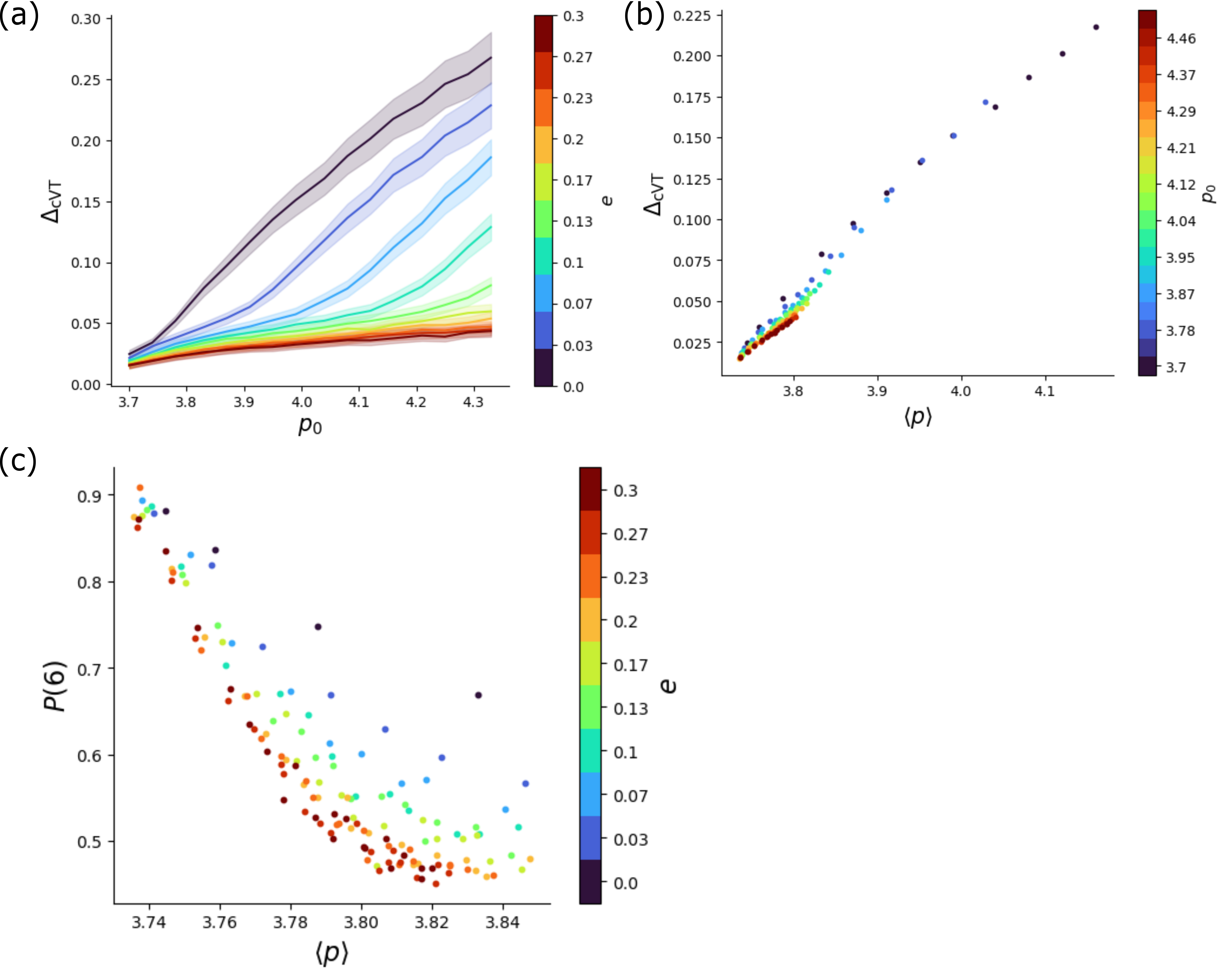}
  \caption{Directly quenching $p_0$ produces the same CVT behaviour as stretch in vertex model cells. For a target quenched shape index $p_0^t$, the degree of quench is denoted by $e$, consistent with the main text and defined from $p^t_0=\frac{p_0}{1+e}$. The degree of quench is discernable from the observable pair (P(6),$\langle p \rangle$) for cells within and about the solid phase.}
  \label{SM: S4}
\end{figure}